\documentclass [12pt]{article}
\usepackage{amsmath, amsthm, amssymb}
\usepackage{bm}
\usepackage{graphicx}
\usepackage{mathptmx}
\usepackage{fullpage}
\usepackage{authblk}
\usepackage{mathrsfs}
\usepackage{enumerate}
\usepackage{makecell}
\usepackage{bibentry, natbib}
\usepackage{float}
\newtheorem{Theorem}{Theorem}
\newtheorem{Lemma}{Lemma}
\newtheorem{Corollary}{Corollary}

\newtheorem{Proposition}{Proposition}
\newtheorem{Example}{Example}
\newtheorem*{Example*}{Example}
\usepackage{tikz}
\usepackage{pgflibraryarrows}
\usepackage{pgflibrarysnakes}
\usepackage{tablefootnote}
\usepackage{booktabs,caption,fixltx2e}
\usepackage[flushleft]{threeparttable}
\usepackage{multirow}
\usepackage{caption}
\emergencystretch=\maxdimen
\title{The Machiavellian frontier of top trading cycles\footnote{The first version: 2021. Chen's research was supported by The National Natural Science Foundation of China No. 72273043 and Jiao's research was supported by The National Natural Science Foundation of China No. 72373095.}}

\author{\small Yajing Chen$^a$\footnote{Corresponding author:
Yajing Chen.  Postal address: 130 Meilong Road, Xuhui District, Shanghai,  200237, China. E-mail address: yajingchen@ecust.edu.cn; yajingchen87@gmail.com},
Zhenhua Jiao$^{b}$,
Chenfeng Zhang$^{a}$, Luosai Zhang$^{c}$
\\
\small $^a$School of Business, East China University of Science and Technology, Shanghai, 200237, China\\
 \small $^b$School of Economics and Trade, Shanghai University of International Business and Economics, Shanghai, 201620, China\\
\small $^c$School of Economics, Shanghai University of Finance and Economics, Shanghai, China
}

\begin{document}
\maketitle

\begin{abstract}
This paper studies the housing market problem introduced by \cite{shapley1974cores}. We probe the Machiavellian frontier of the well-known top trading cycles (TTC) rule by weakening strategy-proofness and providing new characterizations for this rule.  Specifically, our contribution lies in three aspects. First, we weaken the concept of strategy-proofness and introduce a new incentive notion called truncation-invariance, where the truthful preference-reporting assignment cannot be altered by any agent through misreporting a truncation of the true preference at  the assignment produced by the true preference unilaterally.
Second, we characterize the TTC rule by the following three groups of axioms: individual rationality, pair-efficiency, truncation-invariance; individual rationality, Pareto efficiency, truncation-invariance; individual rationality, endowments-swapping-proofness, truncation-invariance.\footnote{Corollary 1 of this paper characterizes the TTC rule through individual rationality, Pareto efficiency and truncation-invariance. Although being an easy implication of the first axioms group, this result still constitutes an important refinement of previous characterizations in the literature. Therefore, we count it as a key result of this paper parallel to the two theorems.} The new characterizations refine several previous results.\footnote{Characterizing by the first group of axioms refines \cite{ma1994strategy}, \cite{chen2021alternative} and \cite{Ekici2023}. Characterizing by the second group of axioms remines \cite{ma1994strategy} and \cite{chen2021alternative}. Characterizing by the third group of axioms refines \cite{fujinaka2018endowments} and \cite{chen2021alternative}.}
Third, we show through examples that the characterization results of \cite{takamiya2001coalition} and \cite{miyagawa2002strategy} can no longer be obtained if strategy-proofness is replaced with truncation-invariance.


\end{abstract}
\section{Introduction}
This paper considers the housing market problem introduced by \cite{shapley1974cores}. In such a problem, there are $n$ agents (traders). Each agent owns a house (object) and has a strict preference over the set of houses. An allocation is a permutation of houses among agents. A rule maps the set of preferences to the set of allocations. \cite{shapley1974cores} not only introduced the classical model of housing market, but also proposed the the top trading cycles (TTC) rule, which selects an allocation via the TTC algorithm (suggested by David Gale). \cite{roth1977weak} later proved that the core (in the sense of weak domination) of the housing market problem coincides with the unique allocation determined by the TTC rule. The TTC rule satisfies nice properties such as individual rationality, Pareto efficiency, strategy-proofness (\cite{roth1982incentive}), and group strategy-proofness (\cite{bird1984group}).

The TTC rule has been characterized in previous works such as \cite{ma1994strategy} by individual rationality, Pareto efficiency, and strategy-proofness\footnote{\cite{svensson1999strategy} gave an alternative proof of \cite{ma1994strategy}'s result. \cite{anno2015short} and \cite{sethuraman2016alternative} gave a shorter proof of \cite{ma1994strategy}'s result.}, \cite{takamiya2001coalition} by strategy-proofness, non-bossiness, individual rationality, and ontoness\footnote{\cite{takamiya2001coalition}'s characterization sharpens \cite{sonmez1996implementation}'s result by weakening Pareto efficiency into ontoness.}, \cite{miyagawa2002strategy} by individual rationality, strategy-proofness, non-bossiness, and anonymity\footnote{In \cite{miyagawa2002strategy}'s results, individual rationality, strategy-proofness, non-bossiness, and anonymity were used to characterize both the TTC rule and the no-trade rule.}, \cite{fujinaka2018endowments} by individual rationality, strategy-proofness and endowments-swapping-proofness, \cite{chen2021alternative} by the following two groups of axioms:  individual rationality, Pareto efficiency, rank monotonicity; individual rationality, endowments-swapping-proofness, rank monotonicity\footnote{\cite{chen2021alternative} refined results of \cite{ma1994strategy} and \cite{fujinaka2018endowments} by using rank monotonicity as a substitute for strategy-proofness.}, and \cite{Ekici2023} by individual rationality, pair-efficiency, and strategy-proofness\footnote{\cite{ekici2024characterizing} provide a short proof of \cite{Ekici2023}'s characterization result. The following works characterize a broader class of rules which include the TTC rule as special cases: \cite{papai2000strategyproof}, \cite{tang2016hierarchical}, \cite{pycia2017incentive}, etc.}.

As shown above, almost all characterization results heavily rely on strategy-proofness. The primary question that we in this paper try to clarify is what is the Machiavellian frontier of the TTC rule under the constraint of axioms combinations such as individual rationality and Pareto efficiency. In other words, we would like to know, to what extent we can relax strategy-proofness while ensuring the uniqueness of TTC allocation or breaking the uniqueness. In order to achieve the above goals, we weaken strategy-proofness based on truncation strategies and use the weaker version to recharacterize the TTC rule.

In the literature, \cite{altuntacs2023some} is closely related to our work. They also probed the Machiavellian frontier of the TTC rule by weakening strategy-proofness. However, their paper is quite different from ours in the following aspects. First, \cite{altuntacs2023some} considered a more general model setting (each agent may be endowed with and consume more than one object) but a more restricted preference domain (lexicographic preference domain). Second, \cite{altuntacs2023some} weakened strategy-proofness by considering several forms of manipulations, but these manipulations are  different from truncation strategies defined in our paper. Third, \cite{altuntacs2023some} probed the Machiavellian frontier between impossibility results and uniqueness of the TTC rule while the current paper goes from uniqueness of the TTC rule to non-uniqueness of it. Therefore, our paper is novel. Specifically, our contribution lies in three aspects.

First, we propose two new axioms based on truncation strategies, namely truncation-invariance and truncation-proofness, to relax the concept of strategy-proofness.
For an agent $i$, we say that a preference $R'_i$ is a truncation of $R_i$ at an object $a$, if any object that is preferred to $a$ under $R'_i$ is ranked in the same place under $R_i$. Truncation-invariance requires that an agent's assignment remains the same if he/she truncates his/her preference at his/her assignment. Truncation-proofness requires that an agent should not be strictly better off if he/she truncates his/her preference at his/her assignment. Truncation-proofness is weaker than truncation-invariance, and truncation-invariance is weaker than both strategy-proofness and rank monotonicity (Proposition 1).

This paper is the first to define truncation-invariance and truncation-proofness, but not the first to define truncation strategies or to weaken strategy-proofness. Truncation strategies were first defined in \cite{chen2017new} and the author used it to define rank monotonicity. Originating from \cite{mongell1991sorority}, there is another larger string of literature studying a different concept of truncation strategies. Typical papers include \cite{roth1991incentives}, \cite{roth1999truncation}, and so on. The same term notwithstanding, the truncation strategies defined in these papers are truncations by moving upwards the outside option if it exists, while truncations defined in the current paper is based on the original allocation of an agent. Moreover, truncation strategies defined in these papers can not be applied to the housing market model.

We motivate restricting each agent's strategy domain to truncations at truthful matching from two perspectives. First, the necessity for restricting manipulations arises from certain inherent limitations in the concept of strategy-proofness.
Recently, there is a string of literature arguing that strategy-proofness lacks empirical support and proposing weakening of this property. Papers such as \cite{charness2009origin} and \cite{esponda2014hypothetical} showed that people have difficulties with hypothetical reasoning even in single-agent decision problems. On seeing this, \cite{troyan2020obvious} proposed the concept of obvious manipulability and non-obvious manipulability, which were used to classify non-strategy-proof rules. Non-obvious manipulability is a weaker requirement than strategy-proofness in the sense of payoff comparison. Unlike \cite{troyan2020obvious}, the current paper weakens strategy-proofness by shrinking the strategy set of agents.  Second, as noted by \cite{altuntacs2023some}, the manipulation based on truncation strategies takes into account evidence from the behavioral economics literature. The literature in this direction has provided strong evidence that in practice an agent often uses heuristics to make decisions (\cite{tversky1974judgment}, \cite{tversky1982judgment}, \cite{gilovich2002heuristics}). \cite{mennle2015power} showed empirical evidence that people may prioritize simple manipulations that are close to their true preferences when they are lying.

Our second contribution consists of new characterizations of the TTC rule. To be specific, we show that the TTC rule in housing markets can be characterized by the following three groups of axioms: individual rationality, pair-efficiency, truncation-invariance (Theorem 1); individual rationality, Pareto efficiency, truncation-invariance (Corollary 1); individual rationality, endowments-swapping-proofness, truncation-invariance (Theorem 2). The new characterizations refine \cite{ma1994strategy}, \cite{fujinaka2018endowments}, \cite{chen2021alternative}, and \cite{Ekici2023} simultaneously. Moreover, these results tell us that we can substantially relax strategy-proofness while ensuring uniqueness of the TTC rule under constraint of axioms combinations like individual rationality and pair-efficiency. 

Finally, some negative results are presented. We show through examples that the characterization result of \cite{takamiya2001coalition}, and \cite{miyagawa2002strategy} can no longer be obtained by weakening strategy-proofness into truncation-invariance. Moreover, we also show through examples that the characterization result of the current paper (Theorems 1, 2 and Corollary 1) can not be further refined by weakening truncation-invariance into truncation-proofness.
The characterization results and negative conclusions of the current paper together help us draw the Machiavellian frontier of the TTC rule.

The main findings of the current paper are summarized in the following two tables.


\begin{center}
{\bf Table 1.} Substitutability of  Truncation-invariance or Truncation-proofnesss for Strategy-proofness in the Literature Results

\smallskip{}

\begin{tabular}{ c|c|cccccccc|ccccccc }
\hline
\multirow{2}*{Literature}& \multirow{2}*{Rules} & \multirow{2}*{IR} & \multirow{2}*{PE} & \multirow{2}*{NB}   & \multirow{2}*{On} & \multirow{2}*{An} & \multirow{2}*{ESP} & \multirow{2}*{PaE}  & \multirow{2}*{SP}  &  \multicolumn{2}{c}{substitutabilities for SP}\\
\cline{11-12}
 &  &  &    &  &  & &  & & & TI & TP\\
\hline
\cite{ma1994strategy}& $\{\tau\}$  & $\checkmark$  & $\checkmark$  &   &   &   &   &   & $\checkmark$  & Yes  &  No    \\
\cite{takamiya2001coalition}& $\{\tau\}$ & $\checkmark$  &   & $\checkmark$  & $\checkmark$  &    &    &    & $\checkmark$  & No  & No     \\
\cite{miyagawa2002strategy}& $\{\tau, NT\}$&$\checkmark$  &  & $\checkmark$  &  & $\checkmark$  &   &   & $\checkmark$  & No  &  No    \\
Fujinaka and & \multirow{2}*{$\{\tau\}$} & \multirow{2}*{$\checkmark$} &   &   &   &   & \multirow{2}*{$\checkmark$}   &  & \multirow{2}*{$\checkmark$}   & \multirow{2}*{Yes} &  \multirow{2}*{No}    \\
Wakayama (2018) &    &   &   &   &   &    &  &    &  &   \\
\cite{Ekici2023}& $\{\tau\}$ & $\checkmark$ &    &   &    &   &    &  $\checkmark$ & $\checkmark$  & Yes  & No    \\
Theorem 1 & $\{\tau\}$& $\checkmark$  &   &   &   &   &   &  $\checkmark$ &   & $\checkmark$  &  No   \\
Corollary 1 & $\{\tau\}$& $\checkmark$  & $\checkmark$   &   &   &   &   &   &   & $\checkmark$  &  No   \\
Theorem 2 & $\{\tau\}$ & $\checkmark$  &  &   &   &   & $\checkmark$  &   &   & $\checkmark$  &  No   \\
\hline
\end{tabular}
\end{center}
The notation ``$\checkmark$'' in a cell means that the axiom is satisfied in the corresponding literature. The notation $\tau$ stands for the TTC rule, and NT stands for the no-trade rule where each agent is assigned his/her endowment. IR, PE, NB, On, An, ESP, PaE, SP, TI, and TP  respectively refer to individual rationality, Pareto efficiency, non-bossiness, ontoness, anonymous, endowments-swapping-proofness, pair-efficiency, strategy-proofness, truncation-invariance and truncation-proofness.

\begin{center}
{\bf Table 2.} Substitutability of  Truncation-invariance or Truncation-proofnesss for Rank Monotonicity in the Literature Results

\smallskip{}

\begin{tabular}{ c|c|cccc|ccccccc }
\hline
\multirow{2}*{Literature}& \multirow{2}*{Rules} & \multirow{2}*{IR} & \multirow{2}*{PE}& \multirow{2}*{ESP} & \multirow{2}*{RM}  &  \multicolumn{2}{c}{substitutabilities for RM}\\
\cline{7-8}
 &  &  &    &  &  &  TI & TP\\
\hline
\cite{chen2021alternative} (Theorem 1)& $\{\tau\}$  & $\checkmark$  & $\checkmark$  &   &    $\checkmark$  & Yes  &  No    \\
\cite{chen2021alternative} (Theorem 2)& $\{\tau\}$ & $\checkmark$  &   & $\checkmark$  & $\checkmark$  & Yes  & No     \\

\hline
\end{tabular}
\end{center}
The notation ``$\checkmark$'' in a cell means that the axiom is satisfied in the corresponding literature. The notation $\tau$ stands for the TTC rule. IR, PE, ESP, RM  respectively refer to individual rationality, Pareto efficiency, endowments-swapping-proofness, rank monotonicity, truncation-invariance and truncation-proofness.


The paper is organized as follows. Section 2 introduces the basic model, relevant axioms and the TTC rule. Section 3 presents the two new axioms based on truncation strategies. Section 4 presents the three new characterizations of the TTC rule. Section 5 discusses some impossibility results. Section 6 concludes.

\section{Preliminaries}

\subsection{Model}

Let $N\equiv\{1,2,\ldots, n\}$ be the set of \textbf{agents} with $n\geq 2$. Let $H\equiv\{h_1,h_2,\ldots, h_n\}$ be the set of \textbf{objects} such that $h_i$ is owned by agent $i\in N$. We call $h_i$ agent $i$'s \textbf{endowment} and also denote it $\omega_i$.
Let $\omega\equiv (\omega_i)_{i\in N} $ denote a profile of endowments. For each $N'\subseteq N$, we use $\omega_{N'}$ to denote the set of endowments belonging to agents in $N'$.
Each agent $i\in N$ has a strict preference relation $P_i$ over $H$. Let $R_i$ be the weak preference relation induced from $P_i$, such that $hR_ih'$ if and only if either $hP_ih{'}$ or $h=h{'}$.
An object $h$ is said to be \textbf{acceptable} to agent $i$ at preference $P_i$ if $h R_i \omega_i$.
Let $\mathcal{P}$ be the set of all strict preferences. Let $P=(P_1, P_2, \ldots, P_n)\in \mathcal{P}^n$ be the preference profile of all agents. Given $i\in N$, $P'_i\in \mathcal{P}$ and $P\in \mathcal{P}^n$, let $-i$ be $N\backslash\{i\}$, and $(P'_i, P_{-i})$ be the preference profile where only $P_i$ is replaced by $P'_i$. For each $i\in N$, $P_i \in \mathcal{P}$, and $h\in H$, let $P_i(h)$ be the preference \textbf{rank} of object $h$ at $P_i$, that is, for each $k\in \{1,2, \ldots, n\}$, $P_i(h)=k$ means that the object $h$ is the $k^{th}$ preferred object in agent $i$'s preference relation $P_i$.
For each $P_i\in \mathcal{P}$ and $h\in H$, let the \textbf{upper contour set} of $P_i$ at $h$ be $U(h,P_i)=\{h{'}\in H: h{'} P_i h\}$.

An \textbf{economy} is a pair $e\equiv (P, \omega)$ consisting of a preference profile and the endowment profile. Let $\mathcal{E}$ be the set of all {possible} economies. An \textbf{allocation} is  a bijection $\mu: N\rightarrow H$, which assigns an object to each agent. For each $N{'}\subseteq N$, we use $\mu_{N{'}}$ to indicate the set of objects assigned to agents in $N{'}$ at allocation $\mu$.
Let $\mathcal{M}$ be the set of all possible allocations.
For any $N{'}\subset N$, an $N{'}-$allocation is a bijection $\mu^{N{'}}: N{'} \rightarrow \omega_{N{'}}$. Thus, an $N-$allocation is an allocation. An $N{'}-$allocation weakly dominates an allocation $\mu$ at an economy $e$ if $\mu^{N{'}}_i R_i \mu_i$ for all $i\in N{'}$ and $\mu^{N{'}}_j P_j \mu_j$ for some $j\in N{'}$. The \textbf{core} of an economy consists of all allocations that are not weakly dominated by any
$N{'}-$allocation.

A \textbf{rule} is a function $\varphi : \mathcal{E} \rightarrow \mathcal{M}$ that selects an allocation for each economy.
For each $N{'}\subseteq N$, we denote the set of objects allocated to agents in $N{'}$ at the economy $e$ under the rule $\varphi$ as $\varphi_{N{'}}(e)$.
Let $\Gamma$ be the set of all possible rules.
\subsection{Axioms}

Pareto efficiency of an allocation requires that no agent can become strictly better off without   other agents becoming worse off by switching to any other allocation.
Formally, a rule $\varphi$ is said to be \textbf{Pareto efficient} if for each $e=(P,\omega)\in \mathcal{E}$, there doesn't exist any $\mu\in \mathcal{M}$ such that $\mu_i R_i \varphi_i(e)$ for each $i\in N$ and $\mu_j P_j \varphi_j(e)$ for some $j\in N$.
We denote the set of all Pareto efficient rules as $\Gamma^{PE}$.

Pair-efficiency rules out efficiency-improving trades between pairs of agents. An allocation $\mu$ is \textbf{pair-efficient} at $e=(P,\omega)\in \mathcal{E}$ if there do not exist $i,j \in N$ such that $\mu_i P_j \mu_j$ and $\mu_j P_i \mu_i$. A rule $\varphi$ is pair-efficient if for each $e=(P,\omega)\in \mathcal{E}$, the allocation $\varphi(e)$ is pair-efficient at $e$.
We denote the set of all pair-efficient rules as $\Gamma^{PaE}$.

The axiom of ontoness says that $\varphi$ is an onto function. Formally, a rule $\varphi$ satisfies \textbf{ontoness} if for $\mu\in \mathcal{M}$, there exists $e=(P,\omega)\in \mathcal{E}$ such that $\varphi(e)=\mu$.

Strategy-proofness of a rule requires that no agent can be strictly better off by misreporting his/her preferences unilaterally. That is, truthtelling is a dominant strategy for each agent under that rule.
Formally, a rule $\varphi$ is \textbf{strategy-proof} if for each $e=(P,\omega)\in \mathcal{E}$, $i\in N$, and $e{'}=((P'_i, P_{-i}), \omega)$, we have $\varphi_i(e)R_i \varphi_i(e{'})$.
{ We denote the set of all  strategy-proof rules as $\Gamma^{SP}$.}

Non-bossiness of a rule requires that: if an agent's assignment remains unaffected when he or she reports a  revised preference, then the assignments of other agents should also stay the same. Formally, a rule $\varphi$ is \textbf{non-bossy} if for each $e=(P,\omega)\in \mathcal{E}$, $i\in N$, and $e{'}=((P'_i, P_{-i}), \omega)\in \mathcal{E}$, if $\varphi_i(e)=\varphi_i(e')$, then $\varphi(e)=\varphi(e')$.

Anonymity of a rule is achieved when agents' assignments are not influenced by their names.
Let $\pi: N\rightarrow N$ be a permutation over $N$ and let $\Pi$ be the set of all permutations. For each $e=(P,\omega)\in \mathcal{E}$ and each $\pi \in \Pi$, let $e^{\pi}=(P^{\pi}, \omega^{\pi})$ be an economy defined as follows: for each $i\in N$, $\omega_i^{\pi}=\omega_{\pi^{-1}(i)}$, and for each $i,j,k\in N$,

\[\omega_j P_i \omega_k \Longleftrightarrow \omega^{\pi}_{\pi(j)} P^{\pi}_{\pi(i)} \omega^{\pi}_{\pi(k)}.\]

\noindent A rule $\varphi$ is \textbf{anonymous} if for each $e=(P,\omega)\in \mathcal{E}$, each $\pi \in \Pi$, and each $i,j \in N$, if $\varphi_i(e)=\omega_j$, then $\varphi_{\pi(i)}(e^{\pi})=\omega^{\pi}_{\pi(j)}$.

Individual rationality of a rule requires that no agent should receive an assignment worse than their initial endowment. Formally,
a rule $\varphi$ is \textbf{individually rational} if for each $e\in \mathcal{E}$ and $i\in N$, we have $\varphi_i(e)R_i \omega_i$.
We denote the set of all individually rational rules as $\Gamma^{IR}$.

Endowments-swapping-proofness of a rule ensures that two agents cannot strictly benefit from swapping their endowments.
Given $e=(P,\omega)\in \mathcal{E}$ and $i,j\in N$, let $e^{ij}=(P, \omega^{ij})$ be the economy such that $\omega^{ij}_i=\omega_j$, $\omega^{ij}_j=\omega_i$, and for each $k\in N\backslash \{i,j\}$, $\omega^{ij}_k=\omega_k$. A rule $\varphi$ is \textbf{endowments-swapping-proof} if there { is} no $e=(P,\omega)\in \mathcal{E}$ and $i,j\in N$ such that $\varphi_i(e^{ij})P_i \varphi_i(e)$ and $\varphi_j(e^{ij})P_j \varphi_j(e)$.
{  We denote the set of all endowments-swapping-proof rules as $\Gamma^{ESP}$.}

To introduce the next monotonicity axiom, we say that $P'_i$ is a \textbf{truncation} of $P_i$ at $h\in H$ if any object that is ranked above $h$ under $P'_i$ has the same rank under $P'_i$ as under $P_i$, that is,
\[
h{'} \, P'_i \, h \Rightarrow P_i(h{'})=P'_i(h{'})\mbox{,  }\forall h{'} \in H.
\]

\noindent Similarly, $P{'}$ is a \textbf{truncation} of $P$ at an allocation $\mu\in \mathcal{M}$ if $P'_i$ is a truncation of $P_i$ at $\mu_i$ for all $i$. Intuitively, if $P{'}$ is a truncation of $P$ at $\mu$, then, some agents have simply moved their assignments under $\mu$ upwards, and the ranks of objects above their assignments under $\mu$ remain the same.

For a rule $\varphi$, the rank monotonicity axiom requires that if $P{'}$ is a truncation of $P$ at $\varphi(P, \omega)$, every agent weakly prefers $\varphi(P{'}, \omega)$ to $\varphi(P, \omega)$ according to $P{'}$.
Formally, a rule satisfies \textbf{rank monotonicity} (\cite{chen2017new}) if for each pair $P , P{'} \in \mathcal{P}$ such that $P{'}$ is a truncation of $P$ at $\varphi(P,\omega)$, $\varphi(P{'},\omega) \,\, R{'} \,\,\varphi(P,\omega)$, which means that $\varphi_i(P{'},\omega) \,\, R'_i \,\,\varphi_i(P,\omega)$ for each $i\in N$. {Note that} rank monotonicity is {a weaker requirement} than weak Maskin monotonicity defined by \cite{kojima2010axioms}, and Maskin monotonicity defined by \cite{maskin1999nash}.

\subsection{The top trading cycles rule}

Let $\tau$ denote the top trading cycles (TTC) rule.
{The TTC rule $\tau$  is defined by providing  each economy $e=(P,\omega)$ an allocation $\tau(e)$ obtained by the following \textbf{top trading cycles (TTC) algorithm}:}

\textbf{\emph{Step 1:}} Each agent points to the owner of his/her first choice object.
There exists at least one cycle and no cycles intersect. Remove all
the cycles and assign each agent in a cycle the object whose owner he/she is pointing to.

~~~~$\vdots$

\textbf{\emph{Step t:}} Each agent points to the owner of his/her first choice object
among the remaining ones. There exists at least one cycle and no cycles
intersect. Remove all the cycles and assign each agent in a cycle the object
whose owner he/she is pointing to.

~~~~$\vdots$

The algorithm terminates in a finite number of steps, when no agent and object remains.

It was shown by \cite{shapley1974cores} that for an economy the core is nonempty.  \cite{roth1977weak} obtained that for any economy, there is a unique allocation in the core which is exactly the allocation produced by the TTC algorithm.

\section{New axioms based on truncation strategies}

We now define two new axioms based on truncation strategies of agents. Truncation-invariance requires that an agent remains the same if he/she truncates his/her preference at his/her assignment.
A rule $\varphi$ satisfies \textbf{truncation-invariance} if for each $e=(P,\omega)\in \mathcal{E}$, $i\in N$, and $e{'}=((P'_i, P_{-i}), \omega)$ such that $P'_i$ is a truncation of $P_i$ at $\varphi_i(P,\omega)$, we have $\varphi_i(e)= \varphi_i(e{'})$.
{We denote the set of all  truncation-invariant rules as $\Gamma^{TI}$.}

Next, we introduce an axiom weaker than truncation-invariance. Truncation-proofness requires that an agent not be strictly better off if he/she truncates his/her preference at his/her assignment. Formally, a rule $\varphi$ satisfies \textbf{truncation-proofness} if for each $e=(P,\omega)\in \mathcal{E}$, $i\in N$, and $e{'}=((P'_i, P_{-i}), \omega)$ such that $P'_i$ is a truncation of $P_i$ at $\varphi_i(P,\omega)$, we have $\varphi_i(e)R_i \varphi_i(e{'})$. Truncation-proofness is weaker than truncation-invariance, and naturally strategy-proofness and rank monotonicity. We denote the set of all truncation-proof rules as $\Gamma^{TP}$.

We now discuss the logical relationship among axioms.

\begin{Proposition}
We have the relationships among  axioms of strategy-proofness, rank
monotonicity, truncation-invariance and truncation-proofness as follows:

\noindent(i) Strategy-proofness implies truncation-invariance, formally, we have $\Gamma^{SP}\subsetneq \Gamma^{TI}$;

\noindent(ii) Rank monotonicity implies truncation-invariance, formally, we have $\Gamma^{RM}\subsetneq \Gamma^{TI}$;

\noindent(iii) Truncation-invariance implies truncation-proofness, formally, we have $\Gamma^{TI}\subsetneq \Gamma^{TP}$.
\end{Proposition}
\begin{proof}
(i) Suppose $\varphi$ satisfies strategy-proofness, but it violates truncation-invariance. Then, there exists $e=(P,\omega)\in \mathcal{E}$, $i\in N$, and $e{'}=((P'_i, P_{-i}), \omega)$ such that $P'_i$ is a truncation of $P_i$ at $\varphi_i(P,\omega)$, and $\varphi_i(e)P_i \varphi_i(e{'})$. Since $P'_i$ is a truncation of $P_i$ at $\varphi_i(e)$, we have $\varphi_i(e) P'_i \varphi_i(e')$, which violates strategy-proofness of $\varphi$.
Note that in the above argument, we consider $e{'}$ as the economy under the true preference, and $e$ as the ``manipulated'' economy, which means that $P'_i$ is the true preference and $P_i$ is the manipulated preference.

Note that the oppositely directed implication of statement (i) is not necessarily true: The immediate acceptance (IA) rule\footnote{See \cite{abdulkadirouglu2003school}.} defined in Example \ref{ex:IA} below satisfies truncation-invariance but violates strategy-proofness.\footnote{From the procedure of the immediate acceptance algorithm, one can easily obtain that the IA rule satisfies truncation-invariance. The IA rule is exactly the so-called Boston mechanism. It is well-known that such mechanism is not strategy-proof.}

(ii) Suppose $\varphi$ satisfies rank monotonicity, but it does not satisfy truncation-invariance. Then, there exists $e=(P,\omega)\in \mathcal{E}$, $i\in N$, and $e{'}=((P'_i, P_{-i}), \omega)$ such that $P'_i$ is a truncation of $P_i$ at $\varphi_i(P,\omega)$, and $\varphi_i(e{'})P'_i \varphi_i(e)$. Since $P'_i$ is a truncation of $P_i$ at $\varphi_i(e)$, $\varphi_i(e{'})P'_i \varphi_i(e)$ implies that $\varphi_i(e{'})P_i  \varphi_i(e)$, which violates rank monotonicity of $\varphi$ because $P_i$ is also a truncation of $P'_i$ at $\varphi_i(e{'})$.

Note that the oppositely directed implication of statement (ii) is necessarily not true: the NRM rule defined in Example \ref{ex:NRM} below satisfies truncation-invariance but violates rank monotonicity.

(iii) It trivially holds by the definition of the two axioms. 
Note that the oppositely directed implication of statement (iii) is not necessarily true: the object-proposing deferred acceptance (OPDA) rule defined in Example \ref{ex:opda} satisfies truncation-proofness but violates truncation-invariance.\footnote{\cite{zhang2024} proved that the OPDA rule satisfies truncation-proofness.}
\end{proof}

\begin{Example}(The immediate acceptance rule, IA)\label{ex:IA}

Let $\mathscr{F}$ denote the set of all bijections from $N$ to $\{1,2, \ldots, |N|\}$. We refer to each of these bijections as a priority order of agents. Each object $h\in H$ has a priority order $\succ_h\in \mathscr{F}$ over agents. Here, $i \succ_h j$ means that agent $i$ has higher priority (or lower priority rank) for object $h$ than agent $j$. The \textbf{priority profile} $\succ \equiv(\succ_h)_{h\in H}$ is a vector of priority orders.

Given $e\equiv (P, \omega)$, choose an arbitrary priority profile, and let $IA(e)$ be the allocation defined by the following immediate acceptance algorithm:

\textbf{\emph{Step 1:}} Each agent applies to his/her favorite object. Each object then accepts the applicant with the highest priority permanently and rejects the other applicants. The agents who are accepted by some objects are removed with their assigned objects.

~~~~$\vdots$

\textbf{\emph{Step t:}} Each remaining agent applies to his/her $t^{th}$ preferred object. Each remaining object then accepts the applicant with the highest priority permanently and rejects the other applicants.

The algorithm terminates in a finite number of  steps, when all agents have been removed.
\end{Example}
\textbf{Remark:} \cite{takamiya2001coalition} introduces individual monotonicity and proves that this axiom is equivalent to strategy-proofness in our model setting. It is then natural to derive that truncation-invariance is also implied by individual monotonicity.

\begin{Example}\label{ex:NRM}

Let $N=\{1,2, \ldots, n\}$ and $H=\{h_1, h_2, \ldots, h_n\}$. For each $i\in N$, let $b_i(e) \in H$ be the top choice of agent $i$ under $e=(P, \omega)$. For each $N^{'}\subset N$, let $e^{N^{'}}$ be the subeconomy induced by removing agents $N\backslash N^{'}$ with their endowments and the preference profile are defined on the remaining agents' endowments. Then let $NRM$ be the rule such that for $e\in \mathcal{E}$, agent $1$ always gets his/her endowment, formally,

$$NRM_1(e)=h_1,$$

\noindent and for each $i\in N\backslash\{1\}$,

$$NRM_i(e)=\left\{
\begin{aligned}
h_i , \mbox{ if }b_1(e)=h_1\\
\tau(e^{N\backslash\{1\}}), \mbox{ if }b_1(e)\neq h_1
\end{aligned}
\right.
$$

\noindent That is, under the NRM rule, agent $1$ is assigned his/her endowment permanently and the other agents' assignments are decided by whether agent $1$'s top choice object is his/her endowment or not. 
This rule is truncation-invariant but not rank monotonic.
\end{Example}

\begin{Example}(The object-proposing deferred acceptance rule, OPDA)\label{ex:opda}

Let $\mathscr{F}$ denote the set of all bijections from $N$ to $\{1,2, \ldots, |N|\}$. We refer to each of these bijections as an order of agents. Each object $h\in H$ has a priority order $\succ_h\in \mathscr{F}$ over agents. Here, $i \succ_h j$ means that agent $i$ has higher priority (or lower priority rank) for object $h$ than agent $j$. The \textbf{priority profile} $\succ \equiv(\succ_h)_{h\in H}$ is a vector of priority orders.

Given an economy $e\equiv (P, \omega)$, choose an arbitrary priority profile, and let $OPDA(e)$ be the allocation defined by the following object-proposing deferred acceptance algorithm:

\textbf{\emph{Step 1:}} Each object applies to the agent who has the highest priority at that object. Each agent
then tentatively accepts the applicant that he/she likes best and rejects the other applicants.

~~~~$\vdots$

\textbf{\emph{Step t:}} Each object that was rejected in the previous step applies to the agent who has the
next priority order. Each agent then tentatively accepts the object that he/she likes best among the new
applicants and the object already tentatively accepted, and rejects the other applicants.

The algorithm terminates in a finite number of steps, when every object has been tentatively
accepted by an agent.  Then every object is assigned to the agent tentatively accepts it.
\end{Example}

\section{New characterizations}

Building on previous characterization results such as \cite{ma1994strategy}, \cite{miyagawa2002strategy}, \cite{fujinaka2018endowments}, \cite{chen2021alternative}, and \cite{Ekici2023}, this paper presents three new characterizations of the TTC rule, which are essentially refinements of the above results.
Our first characterization shows that the TTC rule is the ideal choice when our objective is to find a rule that satisfies individual rationality, pair-efficiency, and truncation-invariance
\footnote{{Thus our Theorem 1 is a refinement of  \cite{Ekici2023}'s Theorem 1.}}.

\begin{Theorem}
A rule satisfies individual rationality, pair-efficiency, and truncation-invariance if and only if it is the TTC rule.
Formally, we have: $\Gamma^{TI}\cap \Gamma^{PaE}\cap \Gamma^{IR}=\{\tau\}$.
\end{Theorem}

The intuition of our proof is as follows: To show the desired result, we introduce the {\bf size} function\footnote{This function was first introduced by \cite{sethuraman2016alternative} and further applied in \cite{ekici2024characterizing}.} for each preference profile $P$, denoted by $s(P)$.  It measures the total number of acceptable objects for agents  at $P$, i.e., $s(P)=\sum_{i \in N} |\{h\in H: hR_i \omega_i\}|$.
Then, we employ the minimal counterexample method:
Suppose there is a rule $\varphi$ that differs from the TTC rule $\tau$ but meets the requirements of individual rationality, pair-efficiency, and truncation-invariance.
We call $P$ a \textbf{conflict profile} if $\varphi(P)\neq \tau(P)$.
Then there is at least one conflict profile.
Among all conflict profiles, let $s(\cdot)$ takes its smallest value at some profile $P^{*}$.
That is
$$P^{*}=\mathop{\arg\min}\limits_{\{P\in\mathcal{P}^n: \varphi(P)\neq \tau(P)\}}{s(P)}.$$
We then modify $P^{*}$ and construct a preference profile $\vec{Q}$ such that pair-efficiency of $\varphi$ contradicts with the truncation-invariance of $\varphi$.
We also provide an alternative proof method which is similar to \cite{Ekici2023}'s proof technique in Appendix A  for readers' reference.

\begin{proof}
It is obvious that $\tau$ satisfies individual rationality, pair-efficiency, and truncation-invariance.

Next, we proceed by contradiction to prove the ``only if'' part.
For convenience of writing, we omit $\omega$ from $e=(P,\omega)$  in this proof where this will not lead any ambiguity.
Let us suppose there is a rule $\varphi$ that differs from the TTC rule $\tau$ but meets the requirements of individual rationality, pair-efficiency, and truncation-invariance.
Then, there exists at least one preference profile $P$ such that the allocations selected by the rule $\varphi$ and the TTC rule are different. i.e., $\varphi(P)\neq \tau(P)$.

Among all the conflict profiles, let $P^{*}$ be the one whose size is smallest. Next, we compare assignments of agents at allocations $\varphi(P^{*})$ and $\tau(P^{*})$ according to $P^{*}$.

Let $\tau(P^{*})$ and $\varphi(P^{*})$ be $x$ and $y$, respectively. Define

\[N_x=\{i\in N: x_i P^{*}_i y_i\},\]

\[N_y=\{i\in N: y_i P^{*}_i x_i\}, \mbox{ and }\]

\[\overline{N}=N\backslash\{ N_x\cup N_y\}=\{i\in N : x_i=y_i\}.\]

Because $x$ is the TTC allocation and thus Pareto efficient, we know that $N_x$ is nonempty while $N_y$ and  $\overline{N}$ are possibly empty.

Next we will show some facts about the shape of $P^{*}$, $x=\tau(P^{*})$ and $y=\varphi(P^{*})$ in the following lemma. For each $h,h^{'}\in H$, we say that $h$ is \textbf{adjacent} to $h^{'}$ at $P_i$ if $P_i(h)=P_i(h^{'})+1$, that is, $h$ is ranked immediately after $h^{'}$ under $P_i$.

\begin{Lemma}
The conflict profile $P^{*}$ with minimal size and the allocations $x=\tau(P^{*})$ and $y=\varphi(P^{*})$ satisfy:

\begin{enumerate}[(1)]

\item for each $i\in N_x$, $y_i$ is adjacent to $x_i$ under $P^{*}_i$ and $\omega_i=y_i$;
\item for each $j\in N_y$, $x_j$ is adjacent to $y_j$ under $P^{*}_j$ and $\omega_j=x_j$;
\item agents in either $N_x$, $N_y$, or $\overline{N}$ are assigned one another's endowments, that is, $x_{N_x}=y_{N_x}=\omega_{N_x}$, $x_{N_y}=y_{N_y}=\omega_{N_y}$, and $x_{\overline{N}}=y_{\overline{N}}=\omega_{\overline{N}}$;
\item the set $N_y$ contains no agent, that is, $N_y=\emptyset$.
\end{enumerate}
\end{Lemma}
\begin{proof}

(1) We will prove this part in two steps by contradiction.

\textbf{Step 1}, $y_i$ is adjacent to $x_i$. Suppose this not true for some $i\in N_x$. Then we can modify agent $i$'s preference order such that $y_i$ is adjacent to $x_i$ under the new preference order. Denote the new preference order by $Q_i$. Then the profile $Q_i$ is a truncation of $P^{*}_i$ at both $x_i$ and $y_i$. Denote the new preference profile by $Q=(Q_i, P^{*}_{-i})$. By truncation-invariance of $\varphi$ and $\tau$, $\varphi(Q)=\varphi(P^{*})\neq \tau(P^{*})=\tau(Q)$. Then $Q$ is a different conflict profile whose size is strictly smaller than $P^{*}$, a contradiction.

\textbf{Step 2}, $\omega_i=y_i$. Suppose this is not true for some $i\in N_x$. Then, by assumption and step 1, $x_iP^{*}_i y_i P^{*}_i \omega_i$. Consider the following preference order of agent $i$, say $\overline{Q}_i$ such that $x_i \overline{Q}_i \omega_i\overline{Q}_i y_i $. That is, change the position of $y_i$ and $\omega_i$ and preserve the relative order of the other objects. Denote the new preference profile by $\overline{Q}=(\overline{Q}_i, P^{*}_{-i})$. By the construction of $\overline{Q}$ we know that the size of $\overline{Q}$ is smaller than $P^{*}$. Since $P^{*}$ is the conflict profile with minimal size, we have $\varphi_i(\overline{Q})=\tau_i(\overline{Q})=x_i$. Now, $P^{*}_i$ is a truncation of $\overline{Q}_i$ at $x_i$, but $\varphi_i(\overline{Q})=x_i \neq y_i=\varphi_i(P^{*})$, which contradicts the assumption that $\varphi$ is truncation-invariant.

(2) The proof is similar to that of part (1) and is thus omitted.

(3) This is a natural corollary of part (1) and (2).

(4) Suppose $N_y\neq \emptyset$. Then, $N_y \cup \overline{N}\neq \emptyset$. By part (3), under the allocation $y$, $N_y \cup \overline{N}$ forms an allocation weakly dominates $x$. This contradicts with the assumption that $x$ is in the core.

\end{proof}

Parts (1), (3) and (4) of Lemma 1  imply the following: There exists a sequence of agents ${N_x}^{*}=\{i_1, i_2, \ldots, i_k, i_{k+1}\}\subseteq N_x, i_{k+1}=i_1$ such that for each $i_s \in \{i_1, i_2, \ldots, i_k\}$, his/her last two acceptable objects are, in order, $x_{i_s}=\omega_{i_{s+1}}$ and $\omega_{i_s}$, as illustrated in the following subprofile of $P^{*}$. Agents in ${N_x}^{*}$ form a cycle during the process of TTC algorithm. Note that $N_x^{*}$ is not necessarily equivalent to $N_x$ because $N_x$ may contain more than one cycles of the TTC algorithm. Here, we assume that $N_x^{*}$ only contains one cycle.

\begin{center}
\begin{tabular}{c|c|c|cc}
$P^{*}_{i_1}$ &$P^{*}_{i_2}$  & $\cdots$  &  $P^{*}_{i_k}$\\
\hline
$\vdots$   &$\vdots$         &  $\vdots$ &  $\vdots$\\
 $\omega_{i_{2}}=\tau_{i_1}(P^{*})$     &$\omega_{i_{3}}=\tau_{i_2}(P^{*})$     &$\vdots$     & $\omega_{i_{1}}=\tau_{i_k}(P^{*})$ \\
$\omega_{i_{1}}=\varphi_{i_1}(P^{*})$      &$\omega_{i_{2}}=\varphi_{i_2}(P^{*})$     &$\vdots$       &  $\omega_{i_{k}}=\varphi_{i_k}(P^{*})$ \\
$\vdots$                    &$\vdots$      &  $\vdots$ &  $\vdots$  \\
\end{tabular}
\end{center}

We also assume that for all the agents in $N_x$, the set ${N_x}^{*}$ are eliminated the earliest by the TTC algorithm under $P^{*}$. Suppose ${N_x}^{*}$ are eliminated in step $t$ of the TTC algorithm. Then, by the definition of TTC algorithm, for each $i_s \in {N_x}^{*}$, the upper contour set of $P^{*}_{i_s}$ at $x_i$, i.e., $U(x_i, P^{*}_{i_s})$ are all objects eliminated during step $1$ to $t-1$ of the TTC algorithm. This implies that  each agent $i_s \in {N_x}^{*}$ finds $\omega_{{N_x}^{*}\backslash\{i_s, i_{s+1}\}}$ unacceptable. (If this is not true, the TTC allocation would have been different.) Moreover, all objects eliminated during step $1$ to $t-1$ of the TTC algorithm are allocated to the same agent under $\varphi$. 

Consider ${N_x}^{*}=\{i_1, i_2, \ldots, i_k\}$. Modify agent $i_k$'s  preference order such that the last $k$ acceptable objects become, in order, $\omega_1, \omega_2, \ldots, \omega_{k-1}, \omega_k$, while the relative rank of the other objects remain the same {as $P^{*}$}. Call the new preference $\vec{Q}_{i_k}$ and the new profile $\vec{Q}=(\vec{Q}_{i_k}, P^{*}_{-i_k})$. Specifically, agents in ${N_x}^{*}$ have the following preferences:

\begin{center}
\begin{tabular}{c|c|c|cc}
$\vec{Q}_{i_1}$ &$\vec{Q}_{i_2}$  & $\cdots$  &  $\vec{Q}_{i_k}$\\
\hline
$\vdots$   &$\vdots$         &  $\vdots$ &  $\vdots$\\
 $\omega_{i_{2}}$     &$\omega_{i_{3}}$     &$\vdots$     & $\omega_{i_{1}}$ \\
$\omega_{i_{1}}$      &$\omega_{i_{2}}$     &$\vdots$       &  $\omega_{i_{2}}$ \\
$\vdots$                    &$\vdots$      &  $\vdots$ &  $\omega_{i_{3}}$  \\
$\vdots$                    &$\vdots$      &  $\vdots$ &  $\vdots$  \\
$\vdots$                    &$\vdots$      &  $\vdots$ &  $\omega_{i_{k}}$  \\
$\vdots$                    &$\vdots$      &  $\vdots$ &  $\vdots$  \\
\end{tabular}
\end{center}

\begin{Lemma}

Under the preference profile $\vec{Q}$, agents in ${N_x}^{*}$ are assigned one another's endowments under $\varphi$, that is, $\varphi_{{N_x}^{*}}(\vec{Q})=\omega_{{N_x}^{*}}$.

\end{Lemma}

\begin{proof}
We assume that for all the agents in $N_x$, the set $N_x^*$ are eliminated the earliest by the TTC algorithm. Suppose $N_x^*$ are eliminated in step $t$ of the TTC algorithm. This implies that all agents eliminated before step $t$ of the TTC algorithm should get the same allocation under $\tau$ with that under $\varphi$. By the procedure of the TTC algorithm, $\tau(P^{*})=\tau(\vec{Q})$ and naturally ${N_x}^{*}$ are also eliminated in step $t$ of TTC under $\vec{Q}$.

When $t=1$, the preference of $i\in N_x^*$ is $P^{*}_i: x_i P^{*}_i \omega_i \ldots$. By individual rationality of $\varphi$, we get the desired conclusion.

When $t\geq2$,
suppose $\varphi_{{N_x}^{*}}(\vec{Q})\neq \omega_{{N_x}^{*}}$.
By the structure of $\vec{Q}$, at least one agent $i^* \in {N_x}^{*}$ gets a strictly better and thus different object under $\varphi$ than under TTC.
By the structure of $\vec{Q}$ again, this object together with agent $i^*$ should be eliminated before step $t$ during the TTC algorithm. We reach a contradiction.
\end{proof}

If $\varphi_{{N_x}^{*}}(\vec{Q})=\omega_{{N_x}^{*}}$, there are two cases.

\textbf{Case 1}, $\varphi_{i_k}(\vec{Q})={\omega_{i_1}}=x_{i_k}$.
Now, $P^{*}_{i_k}$ is a truncation of $\vec{Q}_{i_k}$ at { $\varphi_{i_k}(\vec{Q})=\omega_{i_1}\neq\omega_{i_k}$}.
By truncation-invariance of $\varphi$, $\varphi_{i_k}(\vec{Q})=\varphi_{i_k}(P^{*})=\omega_{i_k}$, a contradiction.

\textbf{Case 2}, $\varphi_{i_k}(\vec{Q})\in \omega_{{N_x}^{*}\backslash \{i_1\}}$. Suppose $\varphi_{i_k}(\vec{Q})=\omega_{i_s}, s\geq 2$. By the structure of $\vec{Q}$, $\varphi_{i_{s-1}}(\vec{Q})=\omega_{i_{s-1}}$ and agent $i_{s-1}$ prefers $\omega_{i_{s}}$. Moreover, $\varphi_{i_k}(\vec{Q})=\omega_{i_s}, s\geq 2$ and agent $i_k$ prefers $\omega_{i_{s-1}}$. This causes a contradiction with pair-efficiency of $\varphi$.

The ``only if'' part of the proof is thus completed.
\end{proof}

\cite{Ekici2023} figures out that within the scope of pair-efficient and individually rational rules, the TTC rule is the unique candidate if we restrict our attention to manipulations through dominant strategies. While Theorem 1 implies that even if we allow only manipulations through truncations, the uniqueness of the TTC rule is still valid.

Since pair-efficiency is a weaker concept than Pareto efficiency and TTC is Pareto efficient, we have the following Corollary implied by Theorem 1. Our second characterization shows that the TTC rule is the ideal choice when our objective is to find a rule that considers both the selection of individual rational and Pareto efficient allocations and the manipulation of agents through truncations.\footnote{Corollary 1 refines Theorem 1 of \cite{ma1994strategy} and Theorem 1 of \cite{chen2021alternative}.}

\begin{Corollary}
A rule satisfies individual rationality, Pareto efficiency, and truncation-invariance if and only if it is the TTC rule.
Formally, we have: $\Gamma^{TI}\cap \Gamma^{PE}\cap \Gamma^{IR}=\{\tau\}$.
\end{Corollary}

According to Proposition 1, strategy-proofness is a strictly stronger requirement than truncation-invariance.
\cite{ma1994strategy} figures out that within the scope of Pareto efficient and individually rational rules, the TTC rule is the unique candidate if we restrict our attention to manipulations through dominant strategies. While Corollary 1 implies that even if we consider only manipulations through truncations, the uniqueness of the TTC rule is still valid.

Our third characterization shows that the TTC rule is ideal if we are seeking for a rule which is individually rational, truncation-invariant and endowments-swapping-proof\footnote{Thus, our theorem 2 is a refinement of \cite{fujinaka2018endowments}'s Theorem 1 and \cite{chen2021alternative}'s Theorem 2.}.

\begin{Theorem}
A rule satisfies individual rationality, endowments-swapping-proofness, and truncation-invariance if and only if it is the TTC rule.
Formally, we have: $\Gamma^{TI}\cap \Gamma^{ESP}\cap \Gamma^{IR}=\{\tau\}$.
\end{Theorem}

The intuition of our proof is as follows: To show the desired result, we introduce the size function for each preference profile and consider the conflict profile $P^{*}$ with minimal size. We then analyze the structure of  $P^{*}$, $\varphi(P^{*})$ and $\tau(P^{*})$ and consider a subset of agents who get strictly better objects under the TTC rule than under the hypothetical rule. We then let agents in this subset exchange endowments until a contradiction is reached. In order to form a comparison with the literature, this article provides an alternative proof method which is similar to \cite{fujinaka2018endowments}'s proof technique in Appendix B  for readers' reference.

\begin{proof}
It is obvious that $\tau$ satisfies individual rationality, endowments-swapping-proofness, and truncation-invariance.

To prove the ``only if'' part,  we use almost the same notation and assumptions as Theorem 1. The only difference is that here we denote an economy by both the endowment and the preference profile. Let $e^{*}=(\omega, P^{*})$. Recall that the proof of Lemma 1 does not rely on the assumption that the hypothetical rule $\varphi$ satisfies pair-efficiency, and only rely on individual rationality and  truncation-invariance of $\varphi$. Thus, Lemma 1 is automatically established here.

Parts (1), (3) and (4) of Lemma 1  imply the following: There exists a sequence of agents ${N_x}^{*}=\{i_1, i_2, \ldots, i_k, i_{k+1}\}\subseteq N_x, i_{k+1}=i_1$ with $|{N_x}^{*}|\geq 2$ such that for each $i_s \in \{i_1, i_2, \ldots, i_k\}$, his/her last two acceptable objects are, in order, $x_{i_s}=\omega_{i_{s+1}}$ and $\omega_{i_s}$. Here $k=|{N_x}^{*}|$. The economy $e^{*}=(\omega, P^{*})$ is described below and $\varphi(e^{*})$  is the allocation marked with boxes. 

\begin{center}
\begin{tabular}{c|c|c|c|c}
$P^{*}_{i_1}$ &$P^{*}_{i_2}$ &$P^{*}_{i_3}$ & $\cdots$  &  $P^{*}_{i_k}$\\
\hline
$\vdots$   &$\vdots$      &  $\vdots$   &  $\vdots$ &  $\vdots$\\
 $\omega_{i_{2}}$     &$\omega_{i_{3}}$  &$\omega_{i_{4}}$   &$\vdots$     & $\omega_{i_{1}}$ \\
$\boxed{\omega_{i_{1}}}$      & $\boxed{\omega_{i_{2}}}$  & $\boxed{\omega_{i_{3}}}$   & $\vdots$       &  $\boxed{\omega_{i_{k}}}$ \\
$\vdots$                    &$\vdots$     &  $\vdots$ &  $\vdots$ &  $\vdots$  \\
\end{tabular}
\end{center}

\begin{center}
\begin{tabular}{c|c|c|c|c}
$\omega_{i_1}$ &$\omega_{i_2}$  &  $\omega_{i_3}$& $\cdots$  &  $\omega_{i_k}$\\
\hline
 $\omega_{i_{1}}$     &$\omega_{i_{2}}$   &$\omega_{i_{3}}$   &$\vdots$     & $\omega_{i_{k}}$ \\
\end{tabular}
\end{center}

We also assume that for all the agents in $N_x$, the set ${N_x}^{*}$ are eliminated the earliest by the TTC algorithm under $P^{*}$. Suppose ${N_x}^{*}$ are eliminated in step $t$ of the TTC algorithm. Then, by the definition of TTC algorithm, for each $i_s \in {N_x}^{*}$, the upper contour set of $P^{*}_{i_s}$ at $x_i$, i.e., $U(x_i, P^{*}_{i_s})$ are all objects eliminated during step $1$ to $t-1$ of the TTC algorithm. This implies that each agent $i_s \in {N_x}^{*}$ finds $\omega_{{N_x}^{*}\backslash\{i_s, i_{s+1}\}}$ unacceptable. (If this is not true, the TTC allocation would have been different.) 

Consider ${N_x}^{*}$. We will next construct a serial of economies by swapping every two adjacent agents' endowments and deduce a contradiction in the end. When $|{N_x}^{*}|=k=2$, agent $i_1$ and $i_2$ like each other's endowment and will become better off simultaneously if they swap endowments. This contradicts endowments-swapping-proofness of $\varphi$.

When $|{N_x}^{*}|=k\geq 3$, we let the adjacent agents exchange their endowments consequently and finally reach a contradiction.

Step 1, let agent $i_1$ and $i_2$ swap their endowments and the preference profile remain the same. Denote the new economy by $e^{i_1i_2}=(\omega^{i_1i_2}, P^{*})$.  The economy $e^{i_1i_2}=(\omega^{i_1i_2}, P^{*})$ is described below. Then agent $i_1$ needs to get $\omega_{i_2}$ by individual rationality of $\varphi$. By endowments-swapping-proofness of $\varphi$, $i_2$ can not get $\omega_{i_3}$. Since $\omega_{i_3}$ has to be assigned to an agent in ${N_x}^{*}$ and only $i_2$ and $i_3$ find it acceptable, we conclude that $i_3$ gets $\omega_{i_3}$. With similar arguments, we know that agents $i_3$ to $i_k$ all get their endowments and agent $i_2$ gets $\omega_{i_1}$. Therefore, $\varphi(e^{i_1i_2})$  is the allocation marked with boxes.

\begin{center}
\begin{tabular}{c|c|c|c|c}
$P^{*}_{i_1}$ &$P^{*}_{i_2}$ &$P^{*}_{i_3}$ & $\cdots$  &  $P^{*}_{i_k}$\\
\hline
$\vdots$   &$\vdots$      &  $\vdots$   &  $\vdots$ &  $\vdots$\\
 $\boxed{\omega_{i_{2}}}$     &$\omega_{i_{3}}$  &$\omega_{i_{4}}$   &$\vdots$     & $\omega_{i_{1}}$ \\
$\omega_{i_{1}}$      &$\omega_{i_{2}}$  &$\boxed{\omega_{i_{3}}}$   &$\vdots$       &  $\boxed{\omega_{i_{k}}}$ \\
$\vdots$                    &$\vdots$     &  $\vdots$ &  $\vdots$ &  $\vdots$  \\
$\vdots$                    &$\boxed{\omega_{i_{1}}}$                    &$\vdots$  &$\vdots$  &  $\vdots$\\
$\vdots$                    &$\vdots$                    &$\vdots$  &$\vdots$                  &$\vdots$ \\
\end{tabular}
\end{center}

\begin{center}
\begin{tabular}{c|c|c|c|c}
$\omega^{i_1i_2}_{i_1}$ &$\omega^{i_1i_2}_{i_2}$  &  $\omega^{i_1i_2}_{i_3}$& $\cdots$  &  $\omega^{i_1i_2}_{i_k}$\\
\hline
 $\omega_{i_{2}}$     &$\omega_{i_{1}}$   &$\omega_{i_{3}}$   &$\vdots$     & $\omega_{i_{k}}$ \\
\end{tabular}
\end{center}

$\vdots$

Step $k-2$, let agent $i_{k-2}$ and $i_{k-1}$ swap their endowments. Denote the new economy by $e^{i_1i_2 \ldots i_{k-1}}=(\omega^{i_1i_2\ldots i_{k-1}}, P^{*})$. The economy $e^{i_1i_2\ldots i_{k-1}}=(\omega^{i_1i_2\ldots i_{k-1}}, P^{*})$ is described below. Then agent $i_1$ to $i_{k-2}$ need to get their endowments by individual rationality of $\varphi$. By endowments-swapping-proofness of $\varphi$, $i_{k-1}$ can not get $\omega_{i_k}$. Since $\omega_{i_k}$ has to be assigned to an agent in ${N_x}^{*}$ and only $i_{k-1}$ and $i_k$ find it acceptable, we conclude that $i_k$ gets $\omega_{i_k}$. Naturally, $i_{k-1}$ gets $\omega_{i_1}$. Therefore, $\varphi(e^{i_1i_2\ldots i_{k-1}})$  is the allocation marked with boxes.

\begin{center}
\begin{tabular}{c|c|c|c|c}
$P^{*}_{i_1}$& $\cdots$  &$P^{*}_{i_{k-2}}$ &$P^{*}_{i_{k-1}}$   &  $P^{*}_{i_k}$\\
\hline
$\vdots$   &$\vdots$      &  $\vdots$   &  $\vdots$ &  $\vdots$\\
 $\boxed{\omega_{i_{2}}}$   &$\vdots$    &$\boxed{\omega_{i_{k-1}}}$  &$\omega_{i_{k}}$      & $\omega_{i_{1}}$ \\
$\omega_{i_{1}}$     &$\vdots$      &$\omega_{i_{k-2}}$  &$\omega_{i_{k-1}}$     &  $\boxed{\omega_{i_{k}}}$ \\
$\vdots$                    &$\vdots$     &  $\vdots$ &  $\vdots$ &  $\vdots$  \\
$\vdots$                                      &$\vdots$  &$\vdots$   &$\boxed{\omega_{i_{1}}}$ &  $\vdots$\\
$\vdots$                    &$\vdots$                    &$\vdots$  &$\vdots$                  &$\vdots$ \\
\end{tabular}
\end{center}

\begin{center}
\begin{tabular}{c|c|c|c|c}
$\omega^{i_1i_2\ldots i_{k-1}}_{i_1}$ & $\cdots$  &$\omega^{i_1i_2\ldots i_{k-1}}_{i_{k-2}}$  &  $\omega^{i_1i_2\ldots i_{k-1}}_{i_{k-1}}$ &  $\omega^{i_1i_2\ldots i_{k-1}}_{i_k}$\\
\hline
 $\omega_{i_{2}}$     &$\vdots$   &$\omega_{i_{k-1}}$   &$\omega_{i_{1}}$     & $\omega_{i_{k}}$ \\
\end{tabular}
\end{center}

Step $k-1$, on the basis of the economy $e^{i_1 i_2 {\ldots} i_{k-1}}$, let agent $i_{k-1}$ and $i_{k}$ swap their endowments. Denote the new economy by $e^{i_1 i_2{\ldots}i_k}=(\omega^{i_1 i_2{\ldots}i_k}, P^{*})$ and it is described below.
Now, the endowments of $i_1, i_2,\ldots, i_k$ are $\omega_{i_2}, \omega_{i_3},\ldots,  \omega_{i_1}$, respectively. Then agent $i_1, i_2, \ldots, i_{k-1}$ need to get their endowments under the new economy $e^{i_1i_2\ldots i_{k}}$ by individual rationality of $\varphi$. Since $\omega_{i_1}$ has to be assigned to an agent in ${N_x}^{*}$, it has to be assigned agent $i_k$. Thus, $\varphi(e^{i_1i_2\ldots i_{k}})$  is the allocation marked with boxes.

\begin{center}
\begin{tabular}{c|c|c|c|c}
$P^{*}_{i_1}$& $\cdots$  &$P^{*}_{i_{k-2}}$ &$P^{*}_{i_{k-1}}$   &  $P^{*}_{i_k}$\\
\hline
$\vdots$   &$\vdots$      &  $\vdots$   &  $\vdots$ &  $\vdots$\\
 $\boxed{\omega_{i_{2}}}$   &$\vdots$    &$\boxed{\omega_{i_{k-1}}}$  &$\boxed{\omega_{i_{k}}}$      & $\boxed{\omega_{i_{1}}}$ \\
$\omega_{i_{1}}$     &$\vdots$      &$\omega_{i_{k-2}}$  &$\omega_{i_{k-1}}$     &  $\omega_{i_{k}}$ \\
$\vdots$                    &$\vdots$     &  $\vdots$ &  $\vdots$ &  $\vdots$  \\
\end{tabular}
\end{center}

\begin{center}
\begin{tabular}{c|c|c|c|c}
$\omega^{i_1i_2\ldots i_{k}}_{i_1}$ & $\cdots$  &$\omega^{i_1i_2\ldots i_{k}}_{i_{k-2}}$  &  $\omega^{i_1i_2\ldots i_{k}}_{i_{k-1}}$ &  $\omega^{i_1i_2\ldots i_{k}}_{i_k}$\\
\hline
 $\omega_{i_{2}}$     &$\vdots$   &$\omega_{i_{k-1}}$   &$\omega_{i_{k}}$     & $\omega_{i_{1}}$ \\
\end{tabular}
\end{center}

Note that agent $i_{k-1}$ and $i_k$ both get a strictly better object by swapping their endowment (from the economy  $e^{i_1 i_2{\ldots}i_{k-1}}=(\omega^{i_1 i_2{\ldots}i_{k-1}}, P^{*})$ to $e^{i_1 i_2{\ldots}i_k}=(\omega^{i_1 i_2{\ldots}i_k}, P^{*})$). This contradicts endowments-swapping-proofness of $\varphi$.
The proof of  the ``only if'' part is thus completed.
\end{proof}

\textbf{Independence of axioms}: We will present the independence of axioms in Theorems 1 (Corollary 1) and 2 simultaneously. The serial dictatorship rule defined in Example \ref{ex:SD} (Borrowed from Example 2 of \cite{fujinaka2018endowments}) violates only individual rationality. The no trade rule defined in Example \ref{ex:NT} (Borrowed from Example 3 of \cite{fujinaka2018endowments}) satisfies both truncation-invariance and individual rationality but violates endowments-swapping-proofness, Pareto efficiency and pair-efficiency. The  $NB^a$ rule defined in Example \ref{ex:NBA} (Borrowed from \cite{fujinaka2018endowments}'s Example 9) violates only truncation-invariance.

\begin{Example}[The serial dictatorial rule, SD]\label{ex:SD}

We now define the serial dictatorial rule, SD. There is a permutation $\pi$ of $N$. The allocation for $e\in  \mathcal{E}$ selected by the SD rule is determined by the following procedure: agent $\pi(1)$ obtains her favorite object in $H$, agent $\pi(2)$ obtains her favorite object in $H \backslash SD_{\pi(1)}(e) $, agent $\pi(3)$ obtains his/her favorite object in $H \backslash \{ SD_{\pi(1)}(e), SD_{\pi(2)}(e) \}$, and so on. It is easy to verify that SD is truncation-invariant, endowments-swapping-proof, and Pareto efficient (and thus pair-efficient), but not individually rational.

\end{Example}

\begin{Example}[The no-trade rule, NT]\label{ex:NT}

The no-trade rule is the rule $NT:\mathcal{E}\rightarrow\mathcal{M}$ such that for each $e=(P,\omega)\in \mathcal{E}$, $NT(e) =\omega$. It is easy to verify that NT is both truncation-invariant and individually rational, but is neither endowments-swapping-proof nor pair-efficient (and thus is not Pareto efficient).

\end{Example}

\begin{Example}\label{ex:NBA}

Let $N=\{1,2,3\}$. Consider $e=(P, \omega)\in \mathcal{E}$  satisfying the following

\begin{center}
\begin{tabular}{ c c c c c c c}
$P_{1}$ & $P_{2}$    & $P_{3}$   &&  $P_3^{'}$  \\
\hline
$h_2$  & $h_1$  &   $h_1$& &$h_1$    \\
$\vdots$   & $h_3$  & $h_2$&& $h_3$   \\
   & $h_2$  & $h_3$& & $h_2$   \\
\end{tabular}
\end{center}

\noindent and $\omega=(h_1,h_2,h_3)$. Let $\mathcal{D} \subset \mathcal{E}$ be the set of economies satisfying conditions stated above. Let $NB^a$ be the rule such that for each $e\in \mathcal{E}$,

$$NB^a(e)=\left\{
\begin{aligned}
(h_2, h_3, h_1), &  &   \mbox{if } e\in \mathcal{D}\\
\tau(e), & & \mbox{otherwise}
\end{aligned}
\right.
$$

\noindent \cite{fujinaka2018endowments} proved that $NB^a$ satisfies  individual rationality, Pareto efficiency, endowments-swapping-proofness, and it is obvious that this rule is different from the TTC rule.

We now prove that this rule violates truncation-invariance. For agent $3$, consider the truncation of his/her preference on $h_1$, denoted by  $P_3^{'}$. We can see that under the new preference profile $(P_1, P_2, P_3^{'})$, agent $3$ will be assigned the object $h_3$, different from his/her previous assignment $h_1$. Therefore, $NB^a$ violates truncation-invariance.

\end{Example}

\section{Discussions}

\subsection{On \cite{takamiya2001coalition}'s characterization}

\cite{takamiya2001coalition} shows that TTC is the unique rule which is strategy-proof, nonbossy, individually rational, and onto. Does this characterization hold even if truncation-invariance is imposed instead of strategy-proofness? The answer is negative. The following  Example \ref{ex:takamiya} demonstrates this point. That is, the $NTM$ rule defined in this example satisfies non-bossiness, individual rationality, ontoness, and truncation-invariance, but is different from the TTC rule.

\begin{Example}\label{ex:takamiya}

Let $N=\{1,2,3\}$. Consider $e=(P, \omega)\in \mathcal{E}$  satisfying the following

\begin{center}
\begin{tabular}{ c c c c c c }
$P_{1}$ & $P_{2}$    & $P_{3}$   \\
\hline
$h_3$  & $h_3$  &   $h_3$  \\
$h_2$  & $h_1$  &   $\vdots$ \\
$h_1$  & $h_2$  &  \\
\end{tabular}
\end{center}

\noindent and $\omega=(h_1,h_2,h_3)$. Let $\mathcal{D} \subset \mathcal{E}$ be the set of economies satisfying conditions stated above. Let $NTM$ be the rule such that for each $e\in \mathcal{E}$,

$$NTM(e)=\left\{
\begin{aligned}
(h_1, h_2, h_3), &  &   \mbox{if } e\in \mathcal{D}\\
\tau(e), & & \mbox{otherwise}
\end{aligned}
\right.
$$

\noindent Obviously $NTM$ satisfies non-bossiness, individual rationality, ontoness, and truncation-invariance.

\end{Example}

\subsection{On \cite{miyagawa2002strategy}'s characterization}

\cite{miyagawa2002strategy} essentially characterizes TTC in terms of strategy-proofness: a rule is individually rational, strategy-proof, non-bossy, and anonymous if and only if it is either TTC or the ``no-trade'' rule. Does this characterization hold even if truncation-invariance is imposed instead of strategy-proofness? The answer is negative. We can check that the $NMY$ rule defined in the following example \ref{ex:miyagawa} (borrowed from \cite{miyagawa2002strategy}'s example 3) is individually rational, non-bossy, anonymous, and truncation-invariant, but it is not equivalent to the TTC rule.

\begin{Example}\label{ex:miyagawa}

Let $b(P_i) \in H$ be the top choice of agent $i$ under $e=(P, \omega)$. Then let $NMY$ be the rule such that for each $e\in \mathcal{E}$,

$$NMY(e)=\left\{
\begin{aligned}
(b(P_1), \ldots, b(P_n)), &  &   \mbox{if } b(P_i)\neq \omega_i \mbox{ for all } i\in N, \mbox{ and }b(P_i)\neq b(P_j) \mbox{ for all } i
= j\\
e, & & \mbox{otherwise}
\end{aligned}
\right.
$$

\noindent That is,the rule selects the endowment allocation except when it is feasible
to give each agent his most preferred object and no agent prefers his endowment.
This rule is not strategy-proof, but it is truncation-invariant since every agent is assigned his/her most preferred object when deviating from the endowment allocation.
\end{Example}

\subsection{On truncation-proofness}

We get both positive and negative results during the above analysis. It is natural then to ask whether we can further refine {Theorem 1, Corollary 1 and Theorem 2} of the current paper by replacing truncation-invariance with truncation-proofness.
Example \ref{ex:NBA} in this paper (borrowed from \cite{fujinaka2018endowments}'s Example 9) shows that the answer is negative , because the $NB^a$ rule defined in this example satisfies individual rationality, Pareto efficiency, pair-efficiency, endowments-swapping-proofness, and truncation-proofness, but is different from the TTC rule.

\begin{Example*} \textbf{6 revisits.}

Let $N=\{1,2,3\}$. Consider $e=(P, \omega)\in \mathcal{E}$  satisfying the following

\begin{center}
\begin{tabular}{ c c c c c c c}
$P_{1}$ & $P_{2}$    & $P_{3}$   && $P_2^{'}$& $P_2^{''}$& $P_3^{'}$  \\
\hline
$h_2$  & $h_1$  &   $h_1$& &$h_3$  & $h_3$  &   $h_1$  \\
$\vdots$   & $h_3$  & $h_2$&& $h_1$  & $h_2$  &   $h_3$ \\
   & $h_2$  & $h_3$& & $h_2$  & $h_1$  &   $h_2$ \\
\end{tabular}
\end{center}

\noindent and $\omega=(h_1,h_2,h_3)$. Let $\mathcal{D} \subset \mathcal{E}$ be the set of economies satisfying conditions stated above. Let $NB^a$ be the rule such that for each $e\in \mathcal{E}$,

$$NB^a(e)=\left\{
\begin{aligned}
(h_2, h_3, h_1), &  &   \mbox{if } e\in \mathcal{D}\\
\tau(e), & & \mbox{otherwise}
\end{aligned}
\right.
$$

\noindent \cite{fujinaka2018endowments} proved that $NB^a$ satisfies  individual rationality, Pareto efficiency, endowments-swapping-proofness, and it is obvious that this rule is different from the TTC rule.

We now prove that this rule satsifies truncation-proofness. Only agent $2$ and $3$ need to be considered. For agent $2$, there are two possible truncations at the allocation $(h_2, h_3, h_1)$, and they are denoted $P_2^{'}$ and $P_2^{''}$, respectively. We can see that if agent $2$ changes his/her preference to $P_2^{'}$ or $P_2^{''}$ while agent $1$ and $3$ do not change their preferences, his/her assignment will not change (agent $2$ still gets object $h_3$), which means that he/she will not be better off. For agent $3$, there is only one possible truncation and it is denoted $P_3^{'}$. We can see that if agent $3$ changes his/her preference to $P_3^{'}$, while agent $1$ and $2$ do not change their preferences, his/her assignment will become worse (change from $h_1$ to $h_3$), which means that he/she will not be better off. Therefore, $NB^a$ satisfies truncation-proofness.

\end{Example*}

\section{Conclusions}

Tables 1 and 2 in Section 1 tell us that under the constraint of different axiom combinations, the Machiavellian frontier of TTC may be different. This paper, through defining axioms much weaker than strategy-proofness, partially draws the boundary line for the Machiavellian frontier of the TTC rule under these axiom combinations. We show to the readers to what extant we can weaken strategy-proofness while keeping or losing the uniqueness of it. The method of finding  Machiavellian frontier of allocation rules is new and can be applied to many other model settings.

This paper formally defines truncation-invariance and truncation-proofness, which are both weaker than strategy-proofness. Both truncation-proofness and truncation-invariance are new in the literature. Thus we set up a benchmark for studying incentive properties of allocation rules. Future research are called for to further study these axioms and apply them to other problems such as school choice and marriage market.

Based on the new axiom that we define, this paper provides three new characterizations of the widely-used TTC rules in addressing housing market problems of \cite{shapley1974cores}. The new characterizations not only refine previous results like \cite{ma1994strategy}, \cite{fujinaka2018endowments}, \cite{chen2021alternative} and \cite{Ekici2023} simultaneously, but also help figuring out the Machiavellian frontier of the TTC rule, which shed new light on the TTC rule.

Moreover, we prove through examples that the characterization result of \cite{takamiya2001coalition}, and \cite{miyagawa2002strategy} can no longer be obtained by weakening strategy-proofness into truncation-invariance. This suggests that strategy-proofness plays a more crucial role in their characterizations. We also prove that if truncation-invariance is replaced by truncation-proofness, all the characterization results mentioned in this paper no longer hold. Future works are needed to investigate the mechanics behind these phenomena.

\section*{Statements and Declarations}

The authors have no competing interests to declare that are relevant to the content of this article.

\bibliography{HM}

\begin{thebibliography}{34}
\providecommand{\natexlab}[1]{#1}

\bibitem[{Abdulkadiro{\u{g}}lu and
  S{\"o}nmez(2003)}]{abdulkadirouglu2003school}
\textsc{Abdulkadiro{\u{g}}lu, A.} and \textsc{S{\"o}nmez, T.} (2003). School
  choice: A mechanism design approach. \textit{American economic review},
  \textbf{93}~(3), 729--747.

\bibitem[{Altunta{\c{s}} \textit{et~al.}(2023)Altunta{\c{s}}, Phan and
  Tamura}]{altuntacs2023some}
\textsc{Altunta{\c{s}}, A.}, \textsc{Phan, W.} and \textsc{Tamura, Y.} (2023).
  Some characterizations of generalized top trading cycles. \textit{Games and
  Economic Behavior}, \textbf{141}, 156--181.

\bibitem[{Anno(2015)}]{anno2015short}
\textsc{Anno, H.} (2015). A short proof for the characterization of the core in
  housing markets. \textit{Economics Letters}, \textbf{126}, 66--67.

\bibitem[{Bird(1984)}]{bird1984group}
\textsc{Bird, C.~G.} (1984). Group incentive compatibility in a market with
  indivisible goods. \textit{Economics Letters}, \textbf{14}~(4), 309--313.

\bibitem[{Charness and Levin(2009)}]{charness2009origin}
\textsc{Charness, G.} and \textsc{Levin, D.} (2009). The origin of the winner's
  curse: {A} laboratory study. \textit{American Economic Journal:
  Microeconomics}, \textbf{1}~(1), 207--36.

\bibitem[{Chen \textit{et~al.}(2024)Chen, Li, Yin, Zhang and Zhou}]{zhang2024}
\textsc{Chen, Q.}, \textsc{Li, Y.}, \textsc{Yin, X.}, \textsc{Zhang, L.} and
  \textsc{Zhou, S.} (2024). Truncation strategies in stable mechanisms.
  \textit{memo}.

\bibitem[{Chen(2017)}]{chen2017new}
\textsc{Chen, Y.} (2017). New axioms for deferred acceptance. \textit{Social
  Choice and Welfare}, \textbf{48}~(2), 393--408.

\bibitem[{Chen and Zhao(2021)}]{chen2021alternative}
\textsc{---} and \textsc{Zhao, F.} (2021). Alternative characterizations of the
  top trading cycles rule in housing markets. \textit{Economics Letters},
  \textbf{201}, 109806.

\bibitem[{Ekici(2023)}]{Ekici2023}
\textsc{Ekici, {\"O}.} (2023). Pair-efficient reallocation of indivisible
  objects. \textit{Theoretical Economics, forthcoming}.

\bibitem[{Ekici and Sethuraman(2024)}]{ekici2024characterizing}
\textsc{Ekici, {\"O}.} and \textsc{Sethuraman, J.} (2024). Characterizing the
  {TTC} rule via pair-efficiency: A short proof. \textit{Economics Letters},
  \textbf{234}, 111459.

\bibitem[{Esponda and Vespa(2014)}]{esponda2014hypothetical}
\textsc{Esponda, I.} and \textsc{Vespa, E.} (2014). Hypothetical thinking and
  information extraction in the laboratory. \textit{American Economic Journal:
  Microeconomics}, \textbf{6}~(4), 180--202.

\bibitem[{Fujinaka and Wakayama(2018)}]{fujinaka2018endowments}
\textsc{Fujinaka, Y.} and \textsc{Wakayama, T.} (2018).
  Endowments-swapping-proof house allocation. \textit{Games and Economic
  Behavior}, \textbf{111}, 187--202.

\bibitem[{Gilovich \textit{et~al.}(2002)Gilovich, Griffin and
  Kahneman}]{gilovich2002heuristics}
\textsc{Gilovich, T.}, \textsc{Griffin, D.} and \textsc{Kahneman, D.} (2002).
  \textit{Heuristics and biases: The psychology of intuitive judgment}.
  Cambridge university press.

\bibitem[{Kojima and Manea(2010)}]{kojima2010axioms}
\textsc{Kojima, F.} and \textsc{Manea, M.} (2010). Axioms for deferred
  acceptance. \textit{Econometrica}, \textbf{78}~(2), 633--653.

\bibitem[{Ma(1994)}]{ma1994strategy}
\textsc{Ma, J.} (1994). Strategy-proofness and the strict core in a market with
  indivisibilities. \textit{International Journal of Game Theory},
  \textbf{23}~(1), 75--83.

\bibitem[{Maskin(1999)}]{maskin1999nash}
\textsc{Maskin, E.} (1999). Nash equilibrium and welfare optimality.
  \textit{The Review of Economic Studies}, \textbf{66}~(1), 23--38.

\bibitem[{Mennle \textit{et~al.}(2015)Mennle, Weiss, Philipp and
  Seuken}]{mennle2015power}
\textsc{Mennle, T.}, \textsc{Weiss, M.}, \textsc{Philipp, B.} and
  \textsc{Seuken, S.} (2015). The power of local manipulation strategies in
  assignment mechanisms. In \textit{IJCAI}, pp. 82--89.

\bibitem[{Miyagawa(2002)}]{miyagawa2002strategy}
\textsc{Miyagawa, E.} (2002). Strategy-proofness and the core in house
  allocation problems. \textit{Games and Economic Behavior}, \textbf{38}~(2),
  347--361.

\bibitem[{Mongell and Roth(1991)}]{mongell1991sorority}
\textsc{Mongell, S.} and \textsc{Roth, A.~E.} (1991). Sorority rush as a
  two-sided matching mechanism. \textit{The American Economic Review}, pp.
  441--464.

\bibitem[{P{\'a}pai(2000)}]{papai2000strategyproof}
\textsc{P{\'a}pai, S.} (2000). Strategyproof assignment by hierarchical
  exchange. \textit{Econometrica}, \textbf{68}~(6), 1403--1433.

\bibitem[{Pycia and {\"U}nver(2017)}]{pycia2017incentive}
\textsc{Pycia, M.} and \textsc{{\"U}nver, M.~U.} (2017). Incentive compatible
  allocation and exchange of discrete resources. \textit{Theoretical
  Economics}, \textbf{12}~(1), 287--329.

\bibitem[{Roth(1982)}]{roth1982incentive}
\textsc{Roth, A.~E.} (1982). Incentive compatibility in a market with
  indivisible goods. \textit{Economics letters}, \textbf{9}~(2), 127--132.

\bibitem[{Roth and Postlewaite(1977)}]{roth1977weak}
\textsc{---} and \textsc{Postlewaite, A.} (1977). Weak versus strong domination
  in a market with indivisible goods. \textit{Journal of Mathematical
  Economics}, \textbf{4}~(2), 131--137.

\bibitem[{Roth and Rothblum(1999)}]{roth1999truncation}
\textsc{---} and \textsc{Rothblum, U.~G.} (1999). Truncation strategies in
  matching markets-in search of advice for participants. \textit{Econometrica},
  \textbf{67}~(1), 21--43.

\bibitem[{Roth and Vate(1991)}]{roth1991incentives}
\textsc{---} and \textsc{Vate, J. H.~V.} (1991). Incentives in two-sided
  matching with random stable mechanisms. \textit{Economic theory},
  \textbf{1}~(1), 31--44.

\bibitem[{Sethuraman(2016)}]{sethuraman2016alternative}
\textsc{Sethuraman, J.} (2016). An alternative proof of a characterization of
  the {TTC} mechanism. \textit{Operations Research Letters}, \textbf{44}~(1),
  107--108.

\bibitem[{Shapley and Scarf(1974)}]{shapley1974cores}
\textsc{Shapley, L.} and \textsc{Scarf, H.} (1974). On cores and
  indivisibility. \textit{Journal of mathematical economics}, \textbf{1}~(1),
  23--37.

\bibitem[{S{\"o}nmez(1996)}]{sonmez1996implementation}
\textsc{S{\"o}nmez, T.} (1996). Implementation in generalized matching
  problems. \textit{Journal of Mathematical Economics}, \textbf{26}~(4),
  429--439.

\bibitem[{Svensson(1999)}]{svensson1999strategy}
\textsc{Svensson, L.-G.} (1999). Strategy-proof allocation of indivisible
  goods. \textit{Social Choice and Welfare}, \textbf{16}~(4), 557--567.

\bibitem[{Takamiya(2001)}]{takamiya2001coalition}
\textsc{Takamiya, K.} (2001). Coalition strategy-proofness and monotonicity in
  {S}hapley-{S}carf housing markets. \textit{Mathematical Social Sciences},
  \textbf{41}~(2), 201--213.

\bibitem[{Tang and Zhang(2016)}]{tang2016hierarchical}
\textsc{Tang, Q.} and \textsc{Zhang, Y.} (2016). Hierarchical exchange rules
  and the core in indivisible objects allocation. \textit{Working paper,
  Shanghai University of Finance and Economics.}

\bibitem[{Troyan and Morrill(2020)}]{troyan2020obvious}
\textsc{Troyan, P.} and \textsc{Morrill, T.} (2020). Obvious manipulations.
  \textit{Journal of Economic Theory}, \textbf{185}, 104970.

\bibitem[{Tversky and Kahneman(1974)}]{tversky1974judgment}
\textsc{Tversky, A.} and \textsc{Kahneman, D.} (1974). Judgment under
  uncertainty: heuristics and biases. \textit{Science}, \textbf{185},
  1124--1131.

\bibitem[{Tversky \textit{et~al.}(1982)Tversky, Kahneman and
  Slovic}]{tversky1982judgment}
\textsc{---}, \textsc{---} and \textsc{Slovic, P.} (1982). \textit{Judgment
  under uncertainty: Heuristics and biases}. Cambridge.

\end{thebibliography}

\bibliographystyle{ecca}

\vskip1cm
\noindent{\Large \bf Appendix A. An alternative proof of Theorem 1}

\begin{proof}
It is obvious that $\tau$ satisfies individual rationality, pair-efficiency, and truncation-invariance.

Next, we proceed by contradiction to prove the ``only if'' part.
We omit $\omega$ from $e=(P,\omega)$ where this will not lead any ambiguity.
Suppose there is a individual rational, pair-efficient, and truncation-invariant rule $\varphi$ that differs from the TTC rule $\tau$.

The intuition of our proof is as follows:
We first redefine the procedure of the TTC rule as in \cite{Ekici2023}, which is outcome-equivalent to the original TTC rule but slightly different in execution.
Then we introduce a TTC-similarity index function $\rho$ to measure the similarity between outcomes recommended by an arbitrary rule and TTC at a given preference profile. 
We denote $\rho(\varphi,P)$ as the TTC-similarity level between the outcomes $\varphi(P)$ and $\tau(P)$. By definition, $\rho(\varphi,P)$ is infinite if the outcomes are the same and finite otherwise.
Then we apply a proof by minimal counterexample method: 
If a rule $\varphi$ satisfying the axioms differs from TTC, $\rho(\varphi,P)$ takes its smallest value for some profile $P$. 
By modifying $P$ and constructing a profile, we demonstrate a contradiction between the pair-efficiency of $\varphi$ and the truncation-invariance of $\varphi$.

We borrow the specification of TTC as in \cite{Ekici2023} that exactly pinpoints the unique cycle executed at each step.
A {\bf cycle} is a sequence where each agent points to his/her favorite object and each object points back to its owner, illustrated as follows:
\[
i_1 \rightarrow \omega_{i_2} \rightarrow i_2 \rightarrow \omega_{i_3} \rightarrow \cdots \rightarrow i_{k-1} \rightarrow \omega_{i_k} \rightarrow i_k \rightarrow \omega_{i_{k+1}} \rightarrow i_{k+1} = i_1
\]
Above, $i_1, i_2, \cdots , i_k$ are distinct agents.
Agent $i_s$ points to object $\omega_{s+1}$ and $\omega_{s+1}$
points to its owner $i_{s+1}$ for $s = 1, \cdots , k$.
The number of agents involved in a cycle $C$ is said to be the {\bf size} of the cycle, denoted by $|C|$.
For instance, the size of the cycle indicated above is $k$.
In comparing two cycles, $C_1$ and $C_2$, $C_1$ is considered {\bf smaller} than $C_2$ if it has fewer elements than $C_2$, or if they have the same number of elements, $C_1$ contains the agent with the smallest index. The smallest cycle in a group is the one that is smaller than all the others.
We assume that at any point in time, TTC proceeds by executing only the smallest cycle.

When running the new specification of  TTC with profile $P$, let $\mathcal{C}_1, \mathcal{C}_2 ,\cdots , \mathcal{C}_t$ denote the sets of cycles that arise at steps $1, 2,\cdots , t$ in order, and let $C_1 \in \mathcal{C}_1, C_2 \in \mathcal{C}_2 ,\cdots , C_t \in \mathcal{C}_t$ be, in order, the cycles that are executed at steps $1, 2,\cdots , t$. We define the {\bf TTC-similarity index function} $\rho$ as follows:
\begin{itemize}
\item At step 1: If $\varphi(P)$ does not execute $C_1$, then $\rho (\varphi, P) = (1, |C_1|)$. If $C_1$ is executed by $\varphi(P)$, we evaluate $\rho (\varphi, P)$ as follows: If TTC terminates at step 1, then $\rho (\varphi, P) = (\infty, \infty)$; otherwise, proceed to step 2.
\item At step $t$: If $\varphi(P)$ does not execute $C_t$, then $\rho (\varphi, P) = (t, |C_t|)$\footnote{When $\rho (\varphi, P) = (t, k)\neq (\infty,\infty)$, it implies that at profile $P$, $\varphi$ initially assigns objects by running TTC with profile $P$ and executing the smallest cycles that arise at steps $1, 2, \cdots, t-1$. However, $\varphi$ deviates from TTC thereafter. Specifically, $\varphi(P)$ does not execute the smallest cycle that arises at step $t$, which is of size $k$.}. If $\varphi(P)$ executes $C_t$, we evaluate $\rho (\varphi, P)$ as follows: If TTC terminates at step $t$, then $\rho (\varphi, P) = (\infty,\infty)$\footnote{If $\rho (\varphi) = (\infty,\infty)$, it indicates that for any profile $P$, $\varphi (P) =\tau(P)$.}; otherwise, proceed to step $t + 1$.
\end{itemize}

We compare the {\bf TTC-similarity levels} using lexicographic order as follows: For $P, P' \in \mathcal{P}^{n}$, if $\rho (\varphi, P) = (x_1, y_1)$ and $\rho (\varphi, P') = (x_2, y_2)$, then $\rho (\varphi, P) \leq \rho (\varphi, P')$ if $x_1 < x_2$, or if $x_1 = x_2$ and $y_1 \leq y_2$. We write $\rho (\varphi, P) < \rho (\varphi, P')$ if $\rho (\varphi, P) \leq \rho (\varphi, P')$ and $\rho (\varphi, P) \neq \rho (\varphi, P')$. Additionally, we define a {\bf TTC-similarity level for a rule $\varphi$ without reference to a preference profile}, denoted by $\rho (\varphi)$. Its value is set equal to the minimum value taken by $\rho (\varphi, P)$ across the set of preference profiles $P \in \mathcal{P}^{n}$, where $\rho (\varphi, P) \leq \rho (\varphi, P')$ for all $P'\in \mathcal{P}^n$.

Since $\varphi$ satisfies the three axioms mentioned above but $\varphi\neq \tau$, we have $\rho (\varphi)\neq (\infty,\infty)$. Let $\rho (\varphi) = (t, k)$ for finite integers $t$ and $k$. Take $P \in \mathcal{P}^{n}$ such that $\rho (\varphi, P) = (t, k)$. This means that $\varphi(P)$ executes cycles $C_1, C_2,\cdots , C_{t-1}$ but not $C_t$.
Let cycle $C_t$ be as follows:
\[
i_1 \rightarrow \omega_{i_2} \rightarrow i_2 \rightarrow \omega_{i_3} \rightarrow . . . \rightarrow i_{k-1} \rightarrow \omega_{i_k} \rightarrow {i_k} \rightarrow \omega_{i_{k+1}} \rightarrow i_{k+1} = i_1.
\]
Let $S = {i_1, i_2,\cdots , i_k}$. Each agent $i_s \in S$ never chooses an object ranked lower than $\omega_{i_{s+1}}$ up to and including step $t$ in TTC. Thus, changing $i_s$'s preference order of objects below $\omega_{i_{s+1}}$ won't alter TTC's execution by this point. Since $\varphi(P)$ skips $C_t$, at least one agent in this cycle isn't assigned his/her pointed object. Suppose instead of $\omega_{i_1}$, $\varphi(P)$ assigns $i_k$ an object ranked lower than $\omega_{i_1}$ at $P_{i_k}$ \footnote{This implys $k \neq 1$. Thus, $k \geq 2$.}.
In the rest of the proof, we'll analyze particular preference relations, illustrated below.
\begin{center}
\begin{tabular}{c|c|cc}
$P_{i_s}$                    &$P_{i_s}^{*}$            &$P_{i_s}^{\dagger}$  &\\
\hline
$\vdots$                      &$\vdots$                   &$\vdots$                    &\\
 $\omega_{i_{s+1}}$   &$\omega_{i_{s+1}}$ &$\omega_{i_{s+1}}$ &\\
$\vdots$                      &$\omega_{i_{s}}$     &$\omega_{i_{k}}$     & \\
$\omega_{i_{s}}$       &$\vdots$                    &$\omega_{i_{s}}$    & \\
$\vdots$                    &$\vdots$                    &$\vdots$                   &  \\
\end{tabular}
\end{center}

Above, $P_{i_s}^{*}$ is obtained from $P_{i_s}$ by moving $\omega_{i_{s}}$ up, positioned right below $\omega_{i_{s+1}}$. 
$P_{i_s}^{\dagger}$ is defined for $s\leq k-2$. It is derived from $P_{i_s}^{*}$ by shifting $\omega_{i_k}$ up, placed right below $\omega_{i_{s+1}}$, and moving $\omega_{i_{s}}$ down, positioned right below $\omega_{i_{k}}$.

\begin{Lemma}\label{claim 1}
Let $P^{1}=(P_{i_1}^{*},P_{i_2}^{*},\cdots,P_{i_k}^{*},P_{I\backslash S})$. For each $i_s\in S$, we have $\varphi_{i_s}(P^{1})=\omega_{i_s}$\footnote{Note that we cannot directly derive this lemma from the proof for \cite{Ekici2023}'s thereom 1. This lemma is strictly stronger than the result in \cite{Ekici2023} since the requirement of truncation-invariance is strictly weaker than strategy-proofness.}.
\end{Lemma}
\begin{proof}
(i) Consider the profile $P^{k} = (P_{i_k}^{*}, P_{I\backslash \{i_k\}})$.
We first claim that $\varphi_{i_k}(P^{k})=\omega_{i_k}$.
Suppose $\varphi_{i_k}(P^{k}) P_{i_k}^{*}\omega_{i_k}$, then $\varphi_{i_k}(P^{k}) R_{i_k} \omega_{i_{k+1}}$. Therefore, $P_{i_k}$ is a truncation of $P_{i_k}^{*}$ at $\varphi_{i_k}(P^{k})$. Since $\varphi$ is truncation-invariant, we have $\varphi_{i_k}(P^{k})=\varphi_{i_k}(P)$. Thus, $\varphi_{i_k}(P) R_{i_k} \omega_{i_{k+1}}$, which contradicts the fact that $\varphi(P)$ assigns $i_k$ an object worse than $\omega_{i_{k+1}}$ at $P$.
Then, also using the fact that $\varphi$ is individually rational, we get $\varphi_{i_k}(P^{k})=\omega_{i_k}$.
When TTC runs with profiles $P^{k}$ and $P$, the same sets of cycles arise, and the same cycles are executed at steps $1, 2,\cdots , t$. Since $\rho (\varphi) = (t, k)$, $\varphi$ executes the same cycles at steps $1, 2,\cdots , {t-1}$. Among the remaining objects after excluding these cycles, $i_{k-1}$'s favorite object at $P_{i_{k-1}}$ is $\omega_{i_k}$. therefore, $\varphi(P^{k})$ assigns $i_{k-1}$ an object worse than $\omega _{i_k}$ at $P_{i_{k-1}}$.

(ii) Consider the profile $P^{k-1} = (P_{i_{k-1}}^{*}, P_{i_k}^{*}, P_{I\backslash\{i_{k-1},i_k\}})$.
By applying a similar argument in (i), it's easy to verify that $\varphi_{i_{k-1}}(P^{k-1})=\omega_{i_{k-1}}$
\footnote{This implies $k\neq 2$. Otherwise, the individual rationality of $\varphi$ implies $\varphi_{i_k}(P^{k-1})=\omega_{i_k}$. However this cannot be true since $i_k$ and $i_{k-1}$ prefer one another's assigned objects to their own assignments while  $\varphi$ is pair-efficient. Thus, $k \geq 3$.}
.
When TTC runs with profiles $P_{k-1}$ and $P$, the same sets of cycles arise, and the same cycles are executed, at steps $1, 2,\cdots , t$.
Since $\rho (\varphi) = (t, k)$, $\varphi$ executes the same cycles at steps $1, 2,\cdots , {t-1}$.
When objects assigned in the these cycles are excluded, among remaining objects, $i_{k-2}$'s favorite object at $P_{i_{k-2}}$ is $\omega_{i_{k-1}}$.
Therefore, $\varphi_{i_{k-1}}$ assigns $i_{k-2}$ an object worse than $\omega_{i_{k-1}}$ at $P_{i_{k-2}}$ .

(iii) Consider the profile $P^{1} = (P_{i_1}^{*}, P_{i_2}^{*},\cdots , P_{i_k}^{*}, P_{I\backslash S})$.
As described above, we can iteratively define the profiles $P^{k-2}, P^{k-3},\cdots , P^{1}$. By similar arguments, for profile $P^{1} = (P_{i_1}^{*}, P_{i_2}^{*},\cdots , P_{i_k}^{*}, P_{I\backslash S})$, we find: $\varphi_{i_1}(P^{1}) = \omega_{i_1}$. When TTC runs with profiles $P^1$ and $P$, the same sets of cycles arise, and the same cycles are executed, at steps $1, 2,\cdots , t$. Then, by individual rationality of $\varphi$, for each $i_s \in  S$, $\varphi_{i_s}(P^{1}) =\omega_{i_s}$.
\end{proof}

\begin{Lemma}
Let $P^{+}= (P_{i_1}^{\dagger}, P_{i_2}^{\dagger},\cdots , P_{i_{k-2}}^{\dagger}, P_{i_{k-1}}^{*}, P_{i_{k}}^{*}, P_{I\backslash S})$. For each $i_s \in S$, we have $\varphi_{i_{s}}(P^{+})= \omega_{i_{s+1}}=\tau_{i_{s}}(P^{+})$.
\end{Lemma}
\begin{proof}
Let $P^{+}= (P_{i_1}^{\dagger}, P_{i_2}^{\dagger},\cdots , P_{i_{k-2}}^{\dagger}, P_{i_{k-1}}^{*}, P_{i_{k}}^{*}, P_{I\backslash S})$, then when TTC runs with profiles $P^{+}$ and $P$, the same sets of cycles arise, and the same cycles are executed, at steps $1, 2,\cdots , t$.
Since $\rho (\varphi) = (t, k)$, $\varphi$ executes the cycles $C_1, C_2,\cdots , C_{t-1}$.
\begin{center}
\begin{tabular}{c|c|c|c|c|c}
$P_{i_1}^{\dagger}$ &$P_{i_2}^{\dagger}$  & $\cdots$ & $P_{i_{k-2}}^{\dagger}$ &$P_{i_{k-1}}^{*}$  & $P_{i_k}^{*}$\\
\hline
$\vdots$                    &$\vdots$                    &$\vdots$  &$\vdots$                                           &$\vdots$         &  $\vdots$ \\
 $\omega_{i_{2}}$     &$\omega_{i_{3}}$     &$\vdots$   &$\omega_{i_{k-1}}$                          & $\omega_{i_{k}}$ & $\omega_{i_{1}}$ \\
$\omega_{i_{k}}$      &$\omega_{i_{k}}$     &$\vdots$   &$\omega_{i_{k}}$     & $\omega_{i_{k-1}}$ & $\omega_{i_{k}}$ \\
$\omega_{i_{1}}$      &$\omega_{i_{2}}$     &$\vdots$   &$\omega_{i_{k-2}}$    &$\vdots$         &  $\vdots$ \\
$\vdots$                    &$\vdots$                    &$\vdots$  &$\vdots$                   &$\vdots$         &  $\vdots$  \\
\end{tabular}
\end{center}

Note that for each $s\leq k-2$, $\varphi_{i_s}(P^{+})\neq \omega_{i_k}$. Otherwise, by the individual rationality of $\varphi$, $\varphi_{i_{k-1}}(P^{+})= \omega_{i_{k-1}}$. By induction, we would have $\varphi_{i_{s+1}}(P^{+})= \omega_{i_{s+1}}$, contradicting the pair-efficiency of $\varphi$.
By applying a similar argument, we have $\varphi_{i_k}(P^{+})\neq \omega_{i_k}$ and
$\varphi_{i_{k-1}}(P^{+})\neq \omega_{i_{k-1}}$.
Hence, $\varphi_{i_{k-1}}(P^{+})=\omega_{i_{k}}$, and by induction based on the individual rationality of $\varphi$, we find $\varphi_{i_{s}}(P^{+})= \omega_{i_{s+1}}=\tau_{i_{s}}(P^{+})$ for each $s\in S$.
\end{proof}

\noindent{\bf Proof for Theorem 1}

\begin{enumerate}
    \item Consider the profile $Q^{k-2} = (P_{i_1}^{\dagger}, P_{i_2}^{\dagger},\cdots , P_{i_{k-3}}^{\dagger}, P_{i_{k-2}}^{*}, P_{i_{k-1}}^{*}, P_{i_{k}}^{*}, P_{I\backslash S})$.
    
    When TTC is run with profiles $Q^{k-2}$ and $P$, the same sets of cycles are generated and executed in steps $1, 2,\cdots , t$. Since $\rho (\varphi) = (t, k)$, $\varphi$ executes cycles $C_1, C_2,\cdots , C_{t-1}$.     
Hence, $P_{i_{k-2}}^{*}$ is a truncation of $P_{i_{k-2}}^{\dagger}$ at $\varphi_{i_{k-2}}(P^{+})=\omega_{i_{k-1}}$. Then,  by the truncation-invariance of $\varphi$, $\varphi_{i_{k-2}}(Q^{k-2})=\varphi_{i_{k-2}}(P^{+})=\omega_{i_{k-1}}$. Further, by the individual rationality of $\varphi$, we have $\varphi_{i_{k-1}}(Q^{k-2})= \omega_{i_{k}}$ and $\varphi_{i_{k}}(Q^{k-2})= \omega_{i_{1}}$. Then, by induction, we obtain $\varphi_{i_{s}}(Q^{k-2})= \omega_{i_{s+1}}=\tau_{i_{s}}(Q^{k-2})$ for each $s\in S$. (This contradicts Lemma \ref{claim 1} when $k=3$.)
    
    \item Consider profile $Q^{k-3} = (P_{i_1}^{\dagger}, P_{i_2}^{\dagger},\cdots , P_{i_{k-4}}^{\dagger}, P_{i_{k-3}}^{*}, P_{i_{k-3}}^{*}, P_{i_{k-1}}^{*}, P_{i_{k}}^{*}, P_{I\backslash S})$.
    
    When TTC is run with profiles $Q^{k-3}$ and $P$, the same sets of cycles are generated and executed in steps $1, 2,\cdots , t$. Since $\rho (\varphi) = (t, k)$, $\varphi$ executes cycles $C_1, C_2,\cdots , C_{t-1}$.
Therefore, $P_{i_{k-3}}^{*}$ is a truncation of $P_{i_{k-3}}^{\dagger}$ at $\varphi_{i_{k-3}}(Q^{k-2})=\omega_{i_{k-2}}$. Then, by the truncation-invariance of $\varphi$, $\varphi_{i_{k-3}}(Q^{k-3})=\varphi_{i_{k-2}}(Q^{k-2})=\omega_{i_{k-2}}$. Further, by the individual rationality of $\varphi$, we have $\varphi_{i_{k-2}}(Q^{k-3})= \omega_{i_{k-1}}$, $\varphi_{i_{k-1}}(Q^{k-3})= \omega_{i_{k}}$, and  $\varphi_{i_{k}}(Q^{k-3})= \omega_{i_{1}}$. Then, by induction, we get $\varphi_{i_{s}}(Q^{k-3})= \omega_{i_{s+1}}=\tau_{i_{s}}(Q^{k-3})$ for each $s\in S$. (This contradicts Lemma \ref{claim 1} when $k=4$.)
    
    \item Consider profile $Q^{1}\equiv P^1 = (P_{i_1}^{*}, P_{i_2}^{*},\cdots , P_{i_k}^{*}, P_{I\backslash S})$.
    
    As described above, we can iteratively define profiles $Q^{k-2}, Q^{k-3},\cdots , Q^{1}$. Using similar arguments, for profile $Q^{1}\equiv P^{1} = (P_{i_1}^{*}, P_{i_2}^{*},\cdots , P_{i_k}^{*}, P_{I\backslash S})$, we find  $\varphi_{i_{s}}(Q^{1})= \omega_{i_{s+1}}=\tau_{i_{s}}(Q^{1})$ for each $s\in S$, which contradicts Lemma \ref{claim 1}.
\end{enumerate}

In summary, we have identified the contradictions required for the proof . 
\end{proof}

\vskip1cm
\noindent{\Large \bf Appendix B. An alternative proof of Theorem 2}

\begin{proof}
It is obvious that $\tau$ satisfies individual rationality, endowments-swapping-proofness, and truncation-invariance.

To prove the ``only if'' part, we begin with a Lemma. Given an economy $e=(P,\omega)\in \mathcal{E}$, let $S=\{i_1, i_2, \ldots, i_K\}$ be a group of agents such that for each $k\in \{0,1,\ldots, K-1\}$ and each $h\in H\backslash \{\omega_{i_{k+1}}\}$, $\omega_{i_{k+1}}P_{i_k}h$, where $i_0\equiv i_K$. Let $\mathcal{L}(e)$ denote the set of such groups. Note that given an economy $e=(P,\omega)\in \mathcal{E}$ and a group $S=\{i_1, i_2, \ldots, i_K\}\in \mathcal{L}(e)$, in step 1 of the TTC algorithm at $e$, a sequence $(\omega_{i_k},i_k)^K_{k=1}$ is a cycle and for each $k\in \{0,1,\ldots, K-1\}$, $\tau_{i_k}(e)=\omega_{i_{k+1}}$.

\begin{Lemma}\label{lem:ESP}
If a rule $\varphi$ satisfies individual rationality, endowments-swapping-proofness, and truncation-invariance, then for each $e \in \mathcal{E}$, each $S\in \mathcal{L}(e)$, and each $i\in S$, $\varphi_i(e)=\tau_i(e)$.
\end{Lemma}

\begin{proof}
We proceed by induction on $|S|$.

\textbf{Base Step:} Let $|S|=1$ and $e=(P,\omega)\in \mathcal{E}$. Since $\{i\}\in \mathcal{L}(e)$, $\omega_i$ is agent $i$'s favorite object at $P_i$. Hence, by individual rationality, $\varphi_i(e)=\tau_i(e)=\omega_i$.

\textbf{Induction Hypothesis:} Suppose the Lemma is true for each $S\in \mathcal{L}(e)$ with $|S|\leq K-1$.

\textbf{Induction Step:} Let $|S|=K$ ($K\geq 2$), $e=(P,\omega)\in \mathcal{E}$, and $S\equiv\{i_1, i_2, \ldots, i_K\}\in \mathcal{L}(e)$ with $i_K=i_0$. Note that by $K\geq 2$, $\omega_{i_{k+1}}\neq \omega_{i_k}$. Let $i_k\in S$. Let $e^{\uparrow}\equiv ((P_{i_k}^{\uparrow}, P_{-i_k}), \omega)$ ,
{ where $P_{i_k}^{\uparrow}$ is $i_k$'s preference such that:
(i) $P_{i_k}^{\uparrow}(\omega_{i_{k+1}})=P_{i_k}^{\uparrow}(\tau_{i_k}(e))=1$;
(ii) $P_{i_k}^{\uparrow}(\omega_{i_k})=2$.

We proceed in two steps.

\textbf{Step 1}: $\varphi_{i_k}(e^{\uparrow})=\omega_{i_{k+1}}$.
By individual rationality, $\varphi_{i_k} (e^{\uparrow})\in \{\omega_{i_{k+1}} ,\omega_{i_k} \}.$
Suppose, by contradiction that $\varphi_{i_{k-1}}(e^{\uparrow}) \neq \omega_{i_k}$,
then $\varphi_{i_k} (e^{\uparrow}) = \omega_{i_k}$.
Since $\omega_{i_k}$ is the favorite object of agent $i_{k-1}$ at $P_{i_{k-1}}$,
we have: (i) $P_{i_{k-1}}(\omega_{i_k})=P_{i_{k-1}}(\tau_{i_{k-1}}(e))=1$;
(ii) $\omega_{i_k}\ P_{i_{k-1}}\ \varphi_{i_{k-1}}(e^{\uparrow})$.

Consider $e^{\uparrow{i_{k-1}}{i_k}} = ((P^{\uparrow}_{i_k},P_{-i_k} ),\omega^{i_{k-1}{i_k}} )$.
Since $\omega^{{i_{k-1}}{i_k}}_{i_{k-1}} = \omega_{i_k}$
and $\omega^{i_{k-1}{i_k}}_{i_k} = \omega_{i_{k-1}}$ ,
the sequences $(\omega_{i_k},{i_{k-1}})$ and $(\omega_{i_1},i_1,\cdots,i_{k-2},\omega{i_{k-1}},i_k,\omega_{i_{k+1}},\cdots,\omega_{i_K} ,i_K )$ are cycles of Step 1 of the TTC algorithm at $e^{\uparrow{i_{k-1}}i_k}$.
Therefore,
$$\{\{i_{k-1}\},\{i_1,\cdots,i_{k-2},{i_k},i_{k+1},\cdots,i_{K} \}\}\subseteq \mathcal{L} (e^{\uparrow{i_{k-1}}{i_k}}).$$
By the induction hypothesis, for each $k'\in\{0, 1,\cdots, K - 1\}$,
$$\varphi_{i_k'}(e^{\uparrow i_{k-1}{i_k}})=\tau_{i_k'}(e^{\uparrow{i_{k-1}}i_k}) = \omega_{i_{k'+1}}.$$
Hence
$$\varphi_{i_k}(e^{\uparrow{i_{k-1}{i_k}}})=\omega_{i_{k+1}}\ P^{\uparrow}_{i_k}\ \omega_{i_k}=\varphi_{i_k} (e^{\uparrow});$$
$$\varphi_{i_{k-1}}(e^{\uparrow{i_{k-1}{i_k}}})=\omega_{i_k}\ P_{i_{k-1}}\ \varphi_{i_{k-1}}(e^{\uparrow}),$$
which contradict endowments-swapping-proofness.
}

\textbf{Step 2}: $\varphi_{i_k}(e)=\omega_{i_{k+1}}$. We can see from the definition that $P_{i_k}$ is a truncation of $P_{i_k}^{\uparrow}$ at $\varphi(e^{\uparrow})$. By truncation-invariance and step 1, i.e., $\varphi_{i_k}(e^{\uparrow})=\omega_{i_{k+1}}$, we have

\[\varphi_{i_k}(e) =\varphi_{i_k}(e^{\uparrow})=\omega_{i_{k+1}}.\]

\end{proof}
To fully prove the ``only if'' part, we need to proceed by induction on the number of agents $n$.
The whole proof can be borrowed from \cite{fujinaka2018endowments}'s Appendix A, and thus is omitted here.
\end{proof}
\end{document}